\documentclass[iop]{emulateapj}
\usepackage{amsmath}
\usepackage[version=3]{mhchem} 
\usepackage{graphicx}
\usepackage{epstopdf}
\def\D{\mathrm{d}}

\usepackage{amssymb}

\shorttitle{GJ 1132b}
\shortauthors{Schaefer et al.}

\begin{document}
\title{Predictions of the atmospheric composition of GJ 1132b}
\author{Laura Schaefer \altaffilmark{1}}
\affil{Harvard-Smithsonian Center for Astrophysics\\
 60 Garden St. \\
 Cambridge, MA 02138}

\author{Robin D. Wordsworth}
\affil{Harvard Paulson School of Engineering and Applied Sciences,\\ 29 Oxford Street, Cambridge, MA 02138}
\affil{Department of Earth and Planetary Sciences,\\ Harvard University, 20 Oxford Street, Cambridge, MA 02138}

\author{Zachory Berta-Thompson}
\affil{MIT\\
	Kavli Institute for Astrophysics and Space Research\\
	77 Massachussetts Ave. Bldg. 37\\
	Cambridge, MA 02139}

\and

\author{Dimitar Sasselov}
\affil{Harvard-Smithsonian Center for Astrophysics\\
 60 Garden St. \\
 Cambridge, MA 02138}

\altaffiltext{1}{lschaefer@cfa.harvard.edu}

\begin{abstract}
GJ 1132 b is a nearby Earth-sized exoplanet transiting an M dwarf, and is amongst the most highly characterizable small exoplanets currently known. In this paper we study the interaction of a magma ocean with a water-rich atmosphere on GJ 1132b and determine that it must have begun with more than 5 wt\% initial water in order to still retain a water-based atmosphere. We also determine the amount of \ce{O2} that can build up in the atmosphere as a result of hydrogen dissociation and loss. We find that the magma ocean absorbs at most $\sim$ 10\% of the \ce{O2} produced, whereas more than 90\% is lost to space through hydrodynamic drag. The most common outcome for GJ 1132 b from our simulations is a tenuous atmosphere dominated by \ce{O2}, although for very large initial water abundances atmospheres with several thousands of bars of \ce{O2} are possible. A substantial steam envelope would indicate either the existence of an earlier \ce{H2} envelope or low XUV flux over the system's lifetime. A steam atmosphere would also imply the continued existence of a magma ocean on GJ 1132 b. Further modeling is needed to study the evolution of \ce{CO2} or \ce{N2}-rich atmospheres on GJ 1132 b. 
\end{abstract}


\section{Introduction} \label{sec:intro}
With the success of the Kepler and K-2 missions and ground-based follow-up efforts of the brightest targets, significant strides have been made in understanding the size and density distribution of planets around other stars \citep[e.g.][]{Burke2015,Dressing2015b}. Planets with radii less than ~1.5 to 1.6 Earth radii and masses less than about 7 Earth masses are universally consistent with a rocky, Earth-like composition \citep{Rogers2015,Weiss2014}. However, most of these likely rocky planets have been found at very close orbital periods and are therefore significantly hotter than the Earth. Some of these planets receive orders of magnitude more stellar insolation than the Earth, and their atmospheres will be sculpted and altered by interactions with the stellar insolation, particularly the high energy extreme ultra-violet (XUV, 1-120 nm) radiation. Therefore models of atmospheric loss and evolution for close-in planets are timely.

There has been substantial work done on atmospheric loss from planets in the solar system, particularly Venus \citep[e.g.,][]{Walker1981,Kasting1983,Zahnle1986,Chassefiere1996,Kulikov2006,Lichtenegger2010,Erkaev2013,Hamano13}. Several recent studies extend this type of modeling to atmospheric loss on habitable zone exoplanets with  \ce{H2O}-rich atmospheres \citep{Wordsworth2013,Wordsworth2014,Ramirez2014,Tian2015,Luger2015}. \citet{Bolmont2016} have modeled water loss from the recently discovered TRAPPIST-1 system of planets around an ultracool dwarf star. Others have also studied whether or not close-in rocky exoplanets could be the residual core remnants of gas giant planets stripped of massive \ce{H2} atmospheres \citep[e.g.,][]{Lammer2009,Lopez2013,Luger2015b,OwenMohanty2016}. Many of the solar system studies have noted that preferential loss of H from steam atmospheres may lead to build up of \ce{O2} in a planet's atmosphere \citep[e.g.,][and references therein]{Kasting1995}. This is particularly a problem for Venus, where minimal \ce{O2} is observed, despite an assumed massive early loss of atmospheric water. \citet{Luger2015} applied this type of model to rocky exoplanets in the habitable zones of M and K dwarf stars, where \ce{O2} may be a biosignature mimic. 

In the present paper, we also study atmospheric loss and oxygen build up, but we extend previous models by including an interior model that allows for uptake of \ce{O2} by the planet's mantle. Our interior model includes both a magma ocean stage, as well as parameterized solid state convection with passive outgassing following solidification. This model is based on magma ocean thermal evolution models long used to study the Solar System terrestrial objects \citep[e.g.,][]{AM85,ET03,Lebrun13,Hamano13}. In comparison, few exoplanet models consider the solid body except as a lower boundary condition for the atmosphere. The present model is an improvement on these treatments and is the first fully coupled model of atmosphere-interior exchange of oxygen. 

We focus on GJ 1132b, a planet only slightly larger than the Earth (M$_{p}$ = 1.62 M$_{\oplus}$, R$_{P}$ = 1.16 R$_{\oplus}$), which was recently discovered by the MEarth ground-based transiting planet survey \citep{Berta15}. GJ 1132 is a nearby M3.5 dwarf (0.181 M$_{\odot}$) located only 12 parsecs away. The planet GJ 1132 b has an orbit of 1.6 days and at 0.0153 AU, receives $\sim$ 19 times more stellar insolation than the Earth and 10 times more than Venus. With a large relative transit depth, GJ 1132 b will be amenable to near-term follow-up both from large ground based telescopes, as well as orbiting observatories like HST and JWST. It is our goal to determine if the planet could have sustained a water or \ce{O2} rich atmosphere over its lifetime. We focus on O and H in order to be able to thoroughly explore the parameter space in a timely manner. Future models may wish to include a more detailed chemistry incorporating carbon and nitrogen-bearing species. 

The magma ocean stage on close-in rocky exoplanets may be extremely long-lived. Observations of these objects may present a means to test magma ocean models which are also used to study processes occuring during Solar System accretion. As such, observations of GJ 1132b and other planets like it may help us improve models for our own Solar System, in particular, models for water and \ce{O2} loss on Venus. 

This paper is organized as follows. Section \ref{sec:escape} discusses our atmospheric escape model and line-by-line climate model. Section \ref{sec:model} describes the planetary interior model and the coupling to the atmospheric model. Section \ref{sec:results} presents results from the coupled model, including the amount of water lost from the planet, the final \ce{O2} abundance in the atmosphere, and the mantle compositon. In section \ref{sec:discussion} we discuss some of the limitations of the model. Finally, in Section \ref{sec:predictions}, we give predictions for the atmospheric composition of GJ 1132 b. 

\section{Atmospheric Escape} \label{sec:escape}

\subsection{Loss of the planet's primordial atmosphere} \label{sec:primordial}

As in the Solar System, atmospheric erosion from young planets around M dwarfs will be driven by a combination of XUV-driven hydrodynamic escape, erosion by coronal mass ejection events (CMEs), blowoff by giant impacts, and a host of more complex processes involving non-thermal effects, ion-pickup and magnetic fields \citep{Khodachenko2007,Lammer2007,Tian2009,Zendejas2010,Vidotto2013,Cohen2014}. The early XUV emission from most M dwarfs is high for an extended period, making XUV-driven hydrodynamic escape one of the most critical effects to model. As it is also more straightforward to calculate escape rates in this case than for many other processes, we focus on it here.

For a planet undergoing XUV-driven hydrodynamic escape, the atmospheric escape flux (kg/m$^2$/s) is approximately given by \citep{Zahnle1990} 
\begin{equation}
\phi  = -\frac{\epsilon F_{XUV}}{4 V_{pot}}\label{eq:Elim}
\end{equation}
where $F_{XUV}$ is the stellar flux in the XUV wavelength range (1-120~nm) and $V_{pot}$ is the gravitational potential at the base of the escaping region. Here we take 
\begin{equation}
V_{pot}=-GM_p/ r_p,
\label{eq:Elim2}
\end{equation}
with $G$ the gravitational constant, and $M_p$ and $r_p$ the estimated planetary radius and mass of GJ 1132 b, respectively (see Table~\ref{tab:params}). $\epsilon$ is an empirical factor that accounts for radiative losses and 3D effects and typically varies between 0.15 and 0.3 \citep{Watson1981,Kasting1983,Chassefiere1996,Tian2009}. Equation~\ref{eq:Elim2} neglects the expansion of the heated upper atmosphere away from the planet's surface, which typically results in a correction factor of up to a few tens of percent.  Equation~\ref{eq:Elim} also assumes that re-emission of absorbed XUV radiation at infrared wavelengths is not effective. This is a reasonable assumption for a hydrogen-dominated upper atmosphere, but not if the upper atmosphere is dominated by a gas with strong vibrational and rotational modes such as \ce{CO2}. We also neglect tidal enhancement of the escape flux \citep{Erkaev2007}, which is likely a much smaller effect than the uncertainty in the XUV flux.

The total mass of atmosphere lost as a fraction of the final planet mass is 
\begin{equation}
\frac{M_{lost}}{M_p} = \frac {4\pi r_p^2} {M_p} \int_0^{t_{now}}\phi(t)\D t = \frac{\pi \epsilon r_p^3}{GM_p^2}\int_0^{t_{now}}F_{XUV}(t) \D t \label{eq:mlost}.
\end{equation}
Here we are assuming $M_{lost}<<M_p$, so that $r_p$ and $M_p$ can be treated as approximately independent of time in eqn (\ref{eq:mlost}).

The present-day XUV flux from GJ 1132 has not yet been measured. However, the star GJ 1214 (0.15 M$_\odot$) is similar in mass to GJ 1132 (0.181 M$_\odot$) and has a similar activity level  \citep{Berta15,Hawley1996}. As such, we use the semi-empirical high-energy spectrum of GJ 1214 constructed by \citet{ParkLoyd2016} as a proxy for that of GJ 1132 (see Fig.~\ref{fig:stellar_flux}). The NUV-to-FUV portion of this spectrum was directly observed with Hubble COS and STIS \citep{France2016}, the EUV was estimated from a model-dependent scaling from the Lyman $\alpha$ emission line \citep{Linsky2014}, and the X-ray from a plasma model matched to an earlier XMM detection of a flare from GJ 1214 \citep{Lalitha2014}. In this spectrum, the XUV flux (1-120nm) represents about 3 $\times~10^{-5}$ of the bolometric flux, with an additional 3 $\times~10^{-5}$ of the bolometric flux contributed by the Lyman $\alpha$ line alone (120-130nm). Based on scaling from GJ 1214, we estimate that GJ 1132 b currently receives about 0.8 W/m$^{2}$ in the XUV. The XUV flux could be at least $3\times$ above or below this value, due both to uncertainties in reconstructing GJ 1214's intrinsic spectrum \citep[see][]{Youngblood2016} and to the unknown extent to which  GJ 1132's high energy behavior tracks that of GJ 1214.

\begin{figure*}
	\begin{center}
		{\includegraphics[scale=0.6]{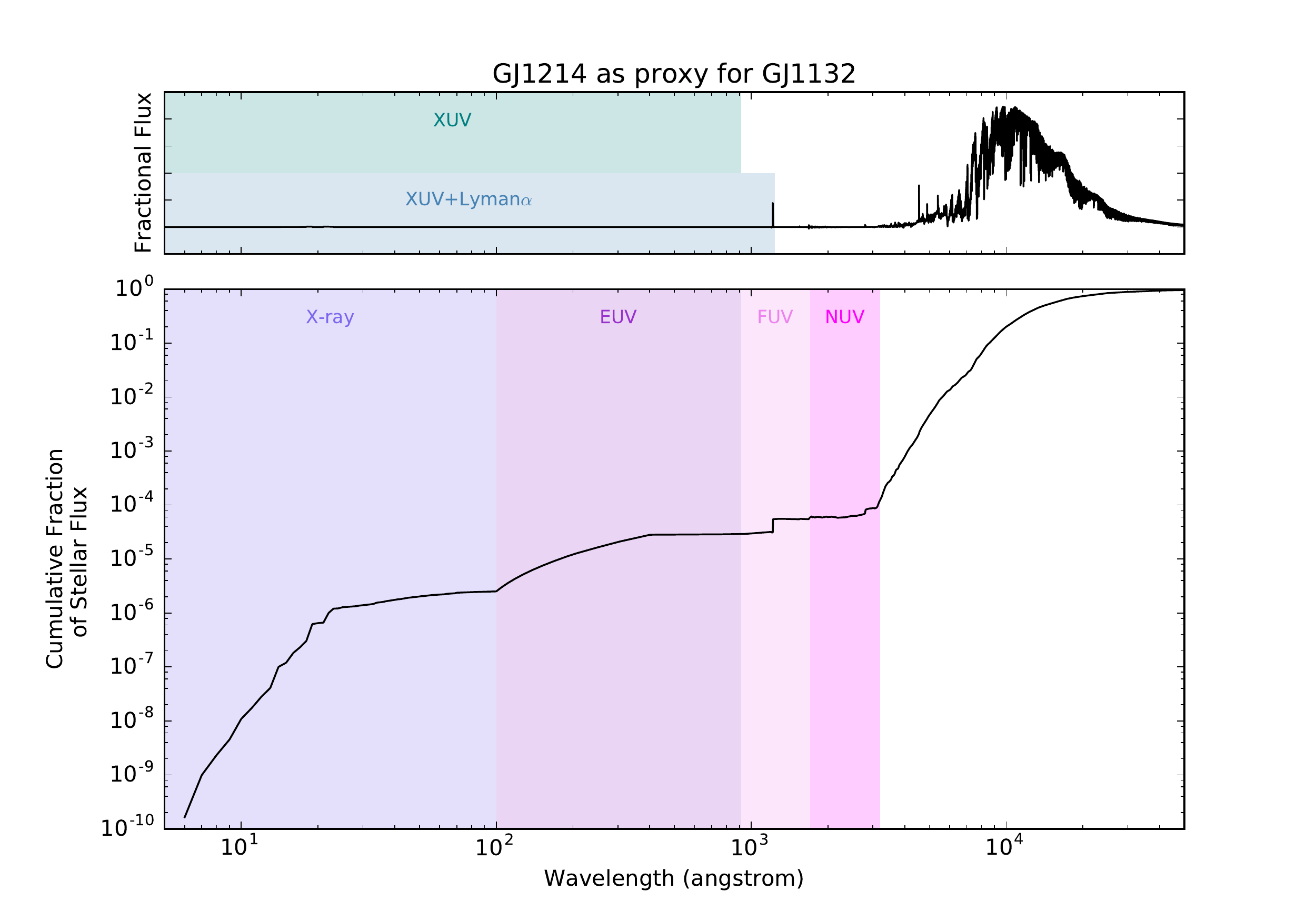}}
	\end{center}
	\caption{Present day spectral energy distribution of GJ 1214 from \citet{ParkLoyd2016}. GJ 1214 serves as a proxy for GJ 1132, for which no measurements of the XUV flux currently exist. The XUV flux for GJ 1214 is approximately $3 \times 10^{-5}$ of the bolometric luminosity, with the Lyman $\alpha$ line containing about an equal amount of flux.}
	\label{fig:stellar_flux}
\end{figure*}

The time evolution of XUV from M dwarfs similar in mass to GJ 1132 is poorly constrained. For main-sequence stars including M dwarfs, observations indicate that time-averaged XUV from the stellar corona for young, active stars saturates at $L_{XUV}=10^{-3}L_{bol}$ \citep{Pizzolato2003,Wright2011}. M dwarfs may stay in this active phase for roughly a gigayear \citep{Shkolnik2014} and then fade to lower XUV flux ratios, although the lower limit for quiescent XUV from old, inactive mid-M dwarfs is just starting to be probed \citep{France2016}. Here, we take two approaches to bracket the range of uncertainty for XUV-driven atmospheric loss. The XUV flux models are shown in Figure \ref{fig:stellar_flux}. XUV flux model A assumes that XUV emissions are $10^{-3}$ times the evolving stellar luminosity \citep{Baraffe2015} and declining as a power law after 1 Gyr with $\epsilon=0.3$. XUV model B assumes that throughout its youth, GJ1132's XUV flux is $10^{-3}$ times the present-day stellar luminosity and zero after 1 Gyr with $\epsilon=0.15$. This brackets the likely present-day XUV flux at 5 Gyr. From eqn. \ref{eq:mlost}, we find ${M_{lost}}/{M_p}=0.142$ for model A. Alternatively, model B yields ${M_{lost}}/{M_p}=0.024$. Hence a very large amount of hydrogen (2\% to 14\% of the total mass) could have been lost from GJ1132b since its formation.

\begin{figure}[h]
	\begin{center}
		{\includegraphics[scale=0.75]{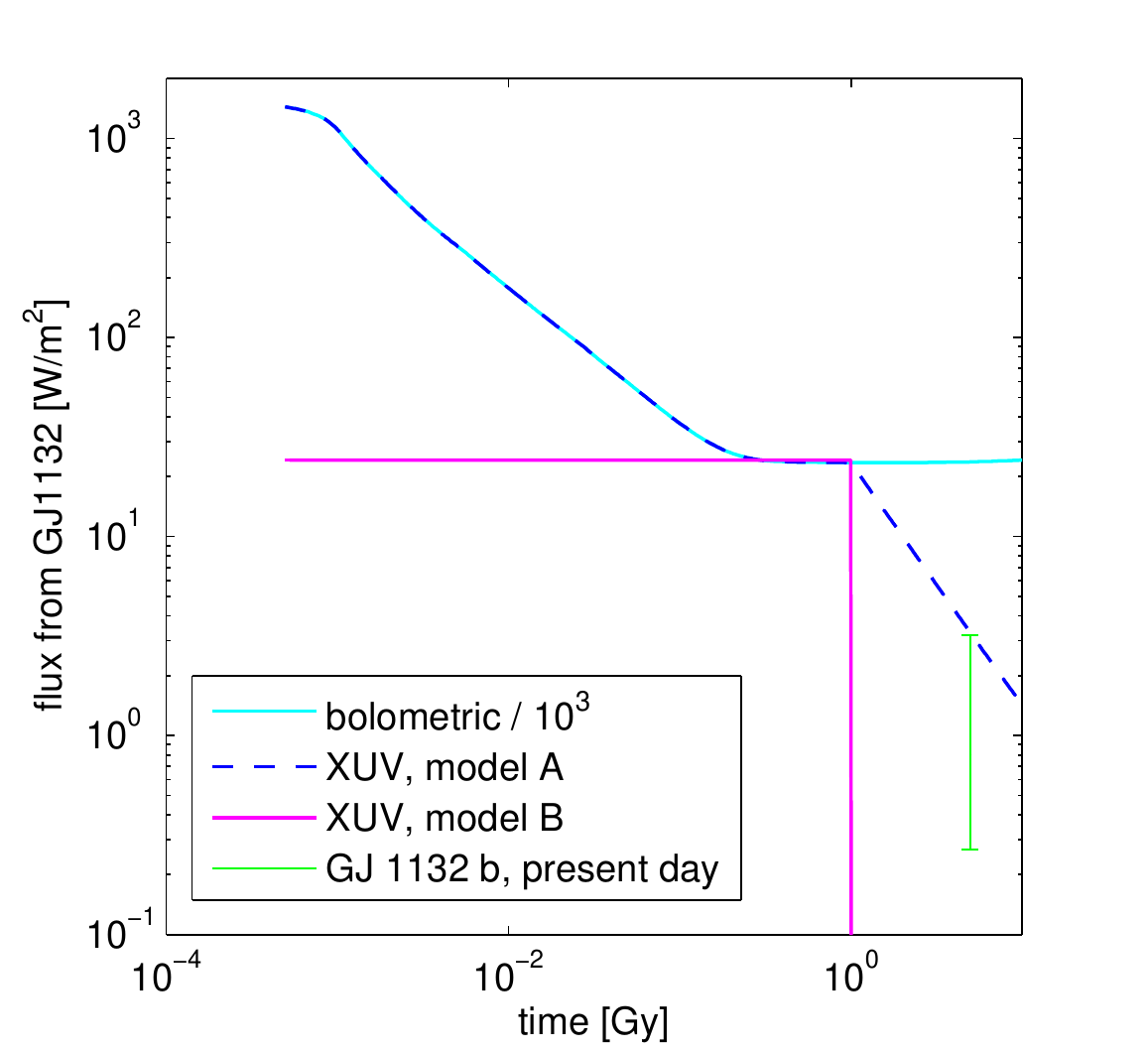}}
	\end{center}
	\caption{Scaled bolometric and XUV flux from GJ 1132 at GJ 1132 b's orbit as a function of time. Bolometric flux was derived by interpolation from the stellar models of \cite{Baraffe2015}. XUV flux was calculated following the method described in the text. The estimated present day XUV flux range for GJ 1132 b is marked with the green bar. }
	\label{fig:stellar_flux}
\end{figure}

\subsection{Drag of heavier species with an escaping hydrogen atmosphere} \label{sec:drag}

Having demonstrated that even a substantial primordial hydrogen envelope could have been lost from GJ 1132 b, we now assess the possibility that the planet has retained an atmosphere of heavier gases. The first thing we need to calculate is the rate at which hydrogen escape would drag away heavier species. 
The flux received by GJ 1132 b places it well within the Kombayashi-Ingersoll limit for the runaway greenhouse\footnote{Given a modern estimate of the Kombayashi-Ingersoll limit of around 280~W/m$^2$ (\cite{Goldblatt2013}; Fig.~\ref{fig:LBL_Goldblatt_intercomp}) a planetary albedo of 0.955 is required for stable surface water on GJ1132b, which is implausibly high for a planet with an atmosphere. Enceladus has an albedo of 0.99 \citep{Verbiscer2007}, but it is airless with a surface composition dominated by fresh water ice.} 
\citep{Kombayashi1967,Ingersoll1969}. If it formed with some water it would hence initially have had an \ce{H2O}-rich upper atmosphere. 

Given an intense early XUV-driven escape regime, the oxygen in this \ce{H2O}, along with other heavy elements such as \ce{C} or \ce{N}, would have been dragged along with the escaping hydrogen. The loss rate of a heavier species in the neutral hydrodynamic escape regime depends on how effectively the hydrogen drags that species with it. Specifically, the number flux $\Phi_2$ of a heavy species 2 [in molecules/m$^2$/s] 
is given by \citep{Hunten1987}
\begin{eqnarray}
\Phi_2 = \left\{
\begin{array}{lr}
\Phi_1 \frac{X_2}{X_1}\frac{\mu_c - \mu_2}{\mu_c - \mu_1}  & : \mu_2<\mu_c\\
0    & : \mu_2>\mu_c
\end{array} \right.
\label{eq:flux_heavy},
\end{eqnarray}
where $X_2$ and $X_1$ and $\mu_1$ and $\mu_2$ are, respectively, the molar concentrations [mol/mol] and molecular masses [amu] of species 1 and 2. The crossover mass $\mu_c$ is defined as 
\begin{equation}
\mu_c = \mu_1 + \frac{k_B T \Phi_1} {b_{12} g X_1 m_p}.\label{eq:crossmass}
\end{equation}
Here $k_B$ is Boltzmann's constant, $T$ is the temperature of the escaping gas, $g$ is gravitational acceleration at the escape radius, $m_p$ is the proton mass and $b_{12}$ is the binary diffusion coefficient for species 1 and 2. For O atoms dragged by H, $b_{12} = 4.8\times 10^{19}T^{0.75}$~m$^{-1}$~s$^{-1}$ \citep{Zahnle1986}. We also define a reference flux 
\begin{equation}
\Phi_{1}^{ref} = \frac{\phi}{\mu_1 m_p} =  \frac{\epsilon F_{XUV}r_p}{4 GM_p\mu_1 m_p}
\label{eq:ref_flux}
\end{equation}
for species 1 in the absence of heavy species \citep{Chassefiere1996}. Note that when $\mu_2>\mu_c$, $\Phi_{1}^{ref}=\Phi_1$ in general. Otherwise, $\mu_{1} \Phi_{1}^{ref} = \mu_{1} \Phi_{1} + \mu_{2} \Phi_{2}$, so that the loss rate depends on the relative abundances of species 1 and 2. In our coupled model, we set $\Phi_{1}$ equal to the diffusion-limited loss rate and $\Phi_{2}$ equal to zero once the abundance of \ce{O2} exceeds that of \ce{H2O}.  Following \citet{Tian2015O2}, we use the composition-dependent loss rates for H and O, rather than the stoichiometric loss rates of \citet{Luger2015}. We discuss the possibility of oxygen-dominated escape from GJ 1132 b in a later section.

Equations (\ref{eq:crossmass}) and (\ref{eq:ref_flux}) can be used to define the critical XUV flux required for drag to occur. Setting $\mu_c = \mu_2$, we can write
\begin{equation}
F^{crit}_{XUV} = \frac{4b_{12}\mathcal V_1^2} { \epsilon k_BTr_p }[\mu_2/\mu_1 - 1] X_1
\end{equation}
with $\mathcal V_1$ the potential energy of one molecule of species 1. For GJ1132b, given O drag by H with $\epsilon=0.3$ and $T=500$~K, $F^{crit}_{XUV} = 0.30$~W/m$^2$. As can be seen from Figure~\ref{fig:stellar_flux}, this is smaller than the estimated XUV flux received by GJ1132b for the first 10~Gy of its lifetime in model A, implying that oxygen will continually be dragged along with escaping hydrogen if an \ce{H2O}-rich atmosphere is present. However, the planet will still oxidize overall as the escape rate of O is less rapid. Whether this oxidation could lead to a detectable atmospheric oxygen signal depends on atmosphere-interior exchange rates, which we address in the Section \ref{sec:model}.

The net buildup rate of O on the planet in the hydrodynamic drag escape regime can be approximated as
\begin{eqnarray}
\Phi^\downarrow_2 &=& b_{12} \frac{n_2}{n_1+n_2}\left(\frac 1 {H_1} - \frac 1 {H_2}\right) \label{eq:Odiff} \\
&=& \frac{b_{12} m_p g}{k_BT}(\mu_1 - \mu_2) X_2
\end{eqnarray}
where the $n_i$ and $H_i$ terms are the molecular number density and individual scale heights of species 1 and 2, respectively. For O diffusing through H following \ce{H2O} photolysis \citep{Luger2015}
\begin{equation}
\Phi^\downarrow_2 \approx -\frac{15b_{12} m_p g}{k_BT}\frac 13 =-\frac{5b_{12} m_p g}{k_BT}.
\end{equation} 
Equation (\ref{eq:Odiff}) can be simply physically interpreted as the diffusion rate of O atoms out of the escaping region back to the lower atmosphere.

\section{Line-by-line Climate Model} \label{sec:linebyline}

The rate at which a planet exchanges volatiles between the atmosphere and interior is a strong function of temperature. In particular, once the surface is hot enough to be in a magma ocean state, the atmosphere and interior will equilibrate on geologically short timescales. For this reason, climate calculations are necessary to assess the increase in surface temperature due to the atmosphere's greenhouse effect.

To calculate surface temperature, we first calculate the outgoing longwave radiation (OLR)  from a pure \ce{H2O} atmosphere using a line-by-line radiative transfer calculation. We integrate the monochromatic equation for upwelling radiative flux per unit wavenumber (W/m$^2$/cm$^{-1}$)
\begin{equation}
\mathcal F_+(\overline \tau_\infty) = \pi B_\nu(T_{surf})e^{-\overline \tau_\infty} + \pi \int_0^{\overline \tau_\infty} B_\nu(\overline \tau) e^{ \overline \tau-\overline \tau_\infty} \D \overline \tau,
\end{equation}
where $T_{surf}$ is surface temperature, $\overline \tau$ is the mean path optical depth at a given wavenumber $ \nu$ and pressure $p$, $\overline \tau_\infty$ is the total optical depth, and $B_\nu$ is the Planck spectral irradiance. Mean path optical depth is defined as 
\begin{equation}
\overline \tau = \frac{\kappa (p_s - p)}{g\overline \mu},
\end{equation}
where $p_s$ is surface pressure, $g$ is surface gravity and $\kappa = \kappa(T,p, \nu)$ is the mass absorption coefficient (m$^2$/kg). In addition, $\overline \mu$ is the mean emission angle cosine, which we take to be a constant 0.5 here. The layer optical depth weighting approach of \cite{Clough1992} is used to ensure accurate model behaviour in high absorption regions of the spectrum. Line absorption coefficients for \ce{H2O} are calculated from the 2010 HITEMP line list \citep{Rothman2013}, with the Voigt function used to describe lineshapes and temperature scaling for the line strengths following standard methods \citep{Rothman1998}. 

The calculation is performed over 30~layers up to a minimum atmospheric pressure of 1~Pa. Spectral calculations were performed from 1~cm$^{-1}$ to 5 times the Wien peak wavenumber of the Planck function at the given surface temperature. We used 5000 points in wavenumber; sensitivity tests indicated that further increases in spectral resolution had an insignificant effect on the integrated OLR.

The temperature profile was assumed to be a dry adiabat from the surface to the tropopause, after which a stratospheric temperature equal to the skin temperature for GJ1132b given a planetary albedo of 0.75 was assumed (344.2~K). Ideal gas behaviour was assumed when calculating the dry adiabat, which is a reasonable approximation for the range of temperatures and pressures studied \citep{Kasting1988,Wordsworth2013}. We accounted for the variation in the specific heat capacity of water vapour as a function of temperature using data from \cite{CRC2000}.

Continuum opacity due to far-wing absorption of strong \ce{H2O} lines and other effects was taken into account using the MT-CKD parametrization \citep{Clough1989}. Outside of the MT-CKD temperature range of validity, continuum absorption was simply set to its value at the maximum temperature given. Spectral lines were truncated at 25~cm$^{-1}$ to avoid double-counting of the continuum absorption. To render the line-by-line calculation more manageable, we also preprocessed the HITEMP-2010 dataset by removing weak lines, which we defined as lines with a reference strength below  $1\times10^{-30}$~cm$^{-1}$ / cm$^2$ molecule$^{-1}$ at 1000~K. This approximation means that we slightly underestimate the atmospheric opacity at the highest temperatures and pressures studied. As the planet's surface is already in a magma ocean state under these conditions, however, this has little effect on  atmospheric evolution.

\begin{figure*}
	\begin{center}
		\includegraphics[scale=0.75,trim = 0.25in 0.4in 0in 0in,clip]{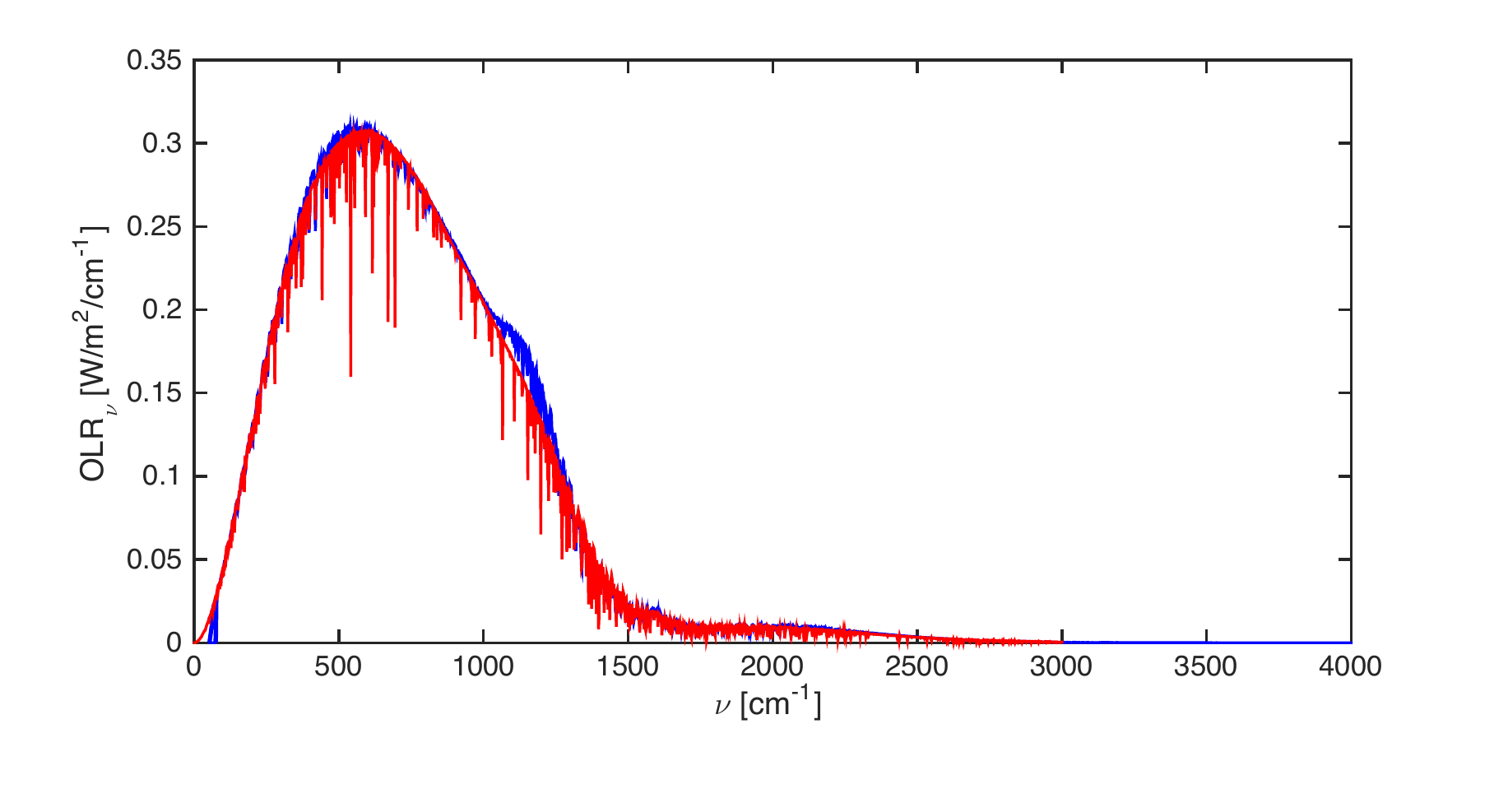}
	\end{center}
	\caption{Outgoing longwave radiation from the line-by-line radiative-convective model (red) vs. results produced using the SMART code detailed in \cite{Goldblatt2013}. In each case the atmospheric composition is 100\% \ce{H2O}, the assumed surface temperature is 300~K and the atmospheric temperature profile follows the \ce{H2O} saturation vapour pressure curve. The spectrally integrated OLR is 281.2~W/m$^2$ and 274.7~W/m$^2$, respectively, for the two cases.}\label{fig:LBL_Goldblatt_intercomp}
\end{figure*}

To validate the code, we first ensured that it reproduced semi-analytic textbook results (Figure~4.5 in \cite{Pierrehumbert2011}). Next, we compared the code output with runaway greenhouse calculations for Earth \citep{Goldblatt2013}. Figure~\ref{fig:LBL_Goldblatt_intercomp} shows the results of this intercomparison. As can be seen, our model agrees closely with published results except in a small region around 1200~cm$^{-1}$, most likely due to slightly differing assumptions for the \ce{H2O} continuum (C. Goldblatt, personal communication). Given the large uncertainties in other parameters for GJ1132b, we decided this agreement was more than sufficient for our purposes.

We calculated the OLR over a range of surface temperatures from 400 to 4000~K and a range of surface pressures from 1~Pa to 1000~bar. At high surface temperatures, the uncertainty in water vapour opacity becomes large due to uncertainty in the scaling of the continuum. However, at these temperatures the planet's surface is in a magma ocean state that permits rapid exchange of oxygen between the atmosphere and interior. Hence this uncertainty should have little effect on our key conclusions.

We calculate the atmospheric heat flux as a balance of the outgoing longwave radiation (OLR) and the absorbed shortwave radiation (ASR). The ASR is given by $(1 - A)F_{stellar}/4$, where $A$ is the planetary albedo and $F_{stellar}$ is the bolometric stellar flux received by the planet, which we derived by interpolating data from \cite{Baraffe2015} to the mass of GJ 1132. The planetary albedo of GJ 1132 b is currently unconstrained, although observations indicate low albedos for planets orbiting M dwarfs in general.  \citet{Demory2014} did a statistical study of Kepler's close-in super-Earths and found a median total albedo of $\sim$ 0.3, although values ranged up to 0.92. Note that the sample of planets studied all had equilibrium temperatures significantly larger than GJ 1132 b, and they may be more representative of bare rocky planets than those with dense atmosphere. Here we take the albedo of Venus, as a representative planet with a thick, hot atmosphere, as our nominal constant value $A=0.75$, but discuss the effect of lower albedos in the results section.

\section{Coupled Atmosphere-Interior Model} \label{sec:model}

We address the atmosphere-interior rates of exchange by coupling our atmospheric model with a magma ocean model, which includes thermal evolution and the exchange of \ce{H2O} and O with the atmosphere. The thermal parameterization combines elements of the work of \citet{Lebrun13}, \citet{ET08}, and \citet{Hamano13}. As in these papers, we assume that solidification of the magma ocean proceeds from the bottom up, due to the fact that mantle adiabats are steeper than the solidus and liquidus curves of silicates. The thermal evolution is governed by two temperatures: the mantle potential temperature, which dictates the degree of melting and convection within the mantle, and the surface temperature, which is governed by heat flux out of the mantle and heat loss from the top of the atmosphere. For most of the duration of the magma ocean phase, these temperatures are the same. However, as the solidification front (the depth at which the mantle adiabat intersects the mantle solidus) moves towards the surface, a thermal boundary layer can develop at the surface, which insulates the mantle from additional heat loss. Following formation of the thermal boundary layer, the model switches to whole mantle solid-state convection as parameterized in \citet{Schaefer15}. The atmosphere is assumed to be composed of \ce{H2O}, H and O gases. \ce{H2O} is the only source of atmospheric opacity and the climate is calculated as discussed above. The composition and thickness of the atmosphere depends on mass exchange with the magma ocean and loss of volatiles due to both atmospheric escape and crystallization into the solid mantle. Following magma ocean solidification, only passive outgassing of \ce{H2O} and atmospheric loss occur. We will discuss each of these aspects in more detail below. 

\subsection{Thermal model} \label{sec: thermal}
The thermal evolution of the magma ocean potential temperature is given by:

\begin{multline}
\frac{4}{3}\pi \rho_{m} c_{p} \left(r^{3}_{p} - r^{3}_{s}\right)\frac{dT_{p}}{dt} = 4\pi r^{2}_{s} \Delta H_{f} \rho_{m} \frac{dr_{s}}{dt} \\
- 4 \pi r^{2}_{p} q_{m} + \frac{4}{3} \pi \rho_{m} Q_{r} \left(r^{3}_{p} - r^{3}_{c}\right)
\end{multline}

where $\rho_{m}$ is the mantle bulk density, $c_{p}$ is the silicate heat capacity ($1.2 \times 10^{3}~J~kg^{-1}~K^{-1}$), $r_{p}$ is the planetary radius, $r_{c}$ is the core radius, $r_{s}$ is the radius of solidification, $\Delta H_{f}$ is the heat of fusion of silicates ($4 \times 10^{5}~J~kg^{-1}$), $q_{m}$ is the mantle heat flux, $Q_{r}$ is the heat generated by radioactive decay. We begin our calculations at $T_p$ = 4000 K, which is hot enough for the magma ocean to extend from the surface to the core-mantle boundary. The heat generated by radioactive decay is limited to the long-lived isotopes \ce{^{40}K}, \ce{^{235,238}U}, and \ce{^{232}Th}. Abundances of these elements are assumed to be the same as for the Earth's mantle, and the parameterization for $Q_{r}$ is the same as that given by \citet{Schaefer15} equation (4). Although we expect GJ 1132 b to have different abundances of the radioactive elements, the results of the magma ocean model are relatively insensitive to them, given the typically short lifetimes of the magma oceans. After solidification, the first term on the RHS disappears and the thermal evolution proceeds as for \citet{Schaefer15}.

The mantle heat flux is parameterized by the mantle Rayleigh number:

\begin{eqnarray}
q_m = \frac{k(T_p - T_{surf})}{l}(\frac{Ra}{Ra_{cr}})^\beta \\
Ra = \frac{\alpha g (T_p - T_{surf})l^3}{\kappa \nu}
\end{eqnarray}

where $k$ is the thermal conductivity (4.2 W m$^{-1}$ K$^{-1}$), the critical Rayleigh number ($1.1 \times 10^{3}$) and the exponent $\beta$ (0.33) are determined from numerical mantle convection simulations, $\alpha$ is the thermal expansion coefficient ($2 \times 10^{-5}$ K$^{-1}$), $\kappa$ is the thermal conductivity ($10^{-6}$ m$^{2}$ s$^{-1}$), and $\nu$ is the kinematic viscosity (m s$^{-2}$). The dynamic viscosity $\eta$ for a silicate liquid is very small, of order 0.01 Pa s. We therefore assume that the liquid portion of the magma ocean is instantaneously well-mixed. We only consider convection within the magma ocean, not the solid mantle, until solidification of the magma ocean has occurred. As partial crystallization proceeds, the viscosity of the magma ocean increases dramatically. The viscosity depends on the melt fraction\footnote{Note that the melt fraction is typically denoted by $\phi$, which we do not use here to avoid confusion with the energy-limited escape flux, see eqn (\ref{eq:Elim2})} $\psi$, which is given by $(T_{p} - T_{solidus})/(T_{liquidus} - T_{solidus})$. We use the same viscosity parameterizations as \citet{Lebrun13}. Below a critical melt fraction ($\psi_c \sim$ 0.4), the viscosity becomes solid-like, where our solid viscosity is given by $\eta = \eta_{0}~exp(-E_{a}/(RT))$, where $\eta_{0} = 3.8 \times 10^{9}~Pa ~s$, $E_{a} = 350$ kJ mole$^{-1}$ and $R$ is the ideal gas constant. 

The radius of solidification is given by the intersection of the mantle adiabat with the mantle solidus. We derive an equation for $(dr_{s}/dt)$ by approximating the adiabat as the first Taylor expansion, and the solidus as a straight line in two sections, from 0 - 100 km, and from 100 km to the core-mantle boundary. The coefficients for the high pressure region are taken from \citet{Hirschmann2000} ($a = 26.53$ K Gpa$^{-1}$, $b$ = 1825 K), and a linear fit is done to the low pressure dry peridotite solidus from that paper ($a = 104.42$ K Gpa$^{-1}$, $b$ = 1420K). The liquidus is assumed to be larger than the solidus by 600 K. The linear parameterization for the solidus leads to a simple and straightforward analytic expression for the radius of solidification, which yields our second differential equation:

\begin{eqnarray}
T_p [1 + \frac{\alpha g}{c_p} (r_p -r_s)] = a g \rho_m (r_p - r_s) + b\\
\frac{dr_s}{dt} = \frac{c_p(b \alpha - a \rho_m c_p)}{g(a \rho_m c_p - \alpha T_p)^2} \frac{dT_p}{dt}
\end{eqnarray}

The surface temperature of the planet is calculated from the heat loss equation for the surface environment, where we make the simplification that the atmosphere and thermal boundary layer are governed by a single average temperature ($T_{surf}$):

\begin{multline}
(c_{p,\ce{H2O}} M_{atm} + c_{p,m} \frac{4}{3} \pi \rho_m (r^3_p - \delta^3)) \frac{dT_{surf}}{dt} \\
= 4 \pi r^2_p (q_m - \mathfrak{F})
\end{multline}

where $\mathfrak{F}$ is the heat flux from the atmosphere, calculated from OLR - ASR (see Sec. \ref{sec:linebyline}), and $\delta$ is the thickness of the thermal boundary layer, which is given by: $\delta = k_m(T_p - T_{surf})/q_m$. The boundary layer develops once the melt fraction at the surface of the magma ocean reaches the critical value, causing the viscosity of the magma ocean to increase dramatically. This is the "mush" stage of \citet{Lebrun13}.

\begin{figure}
	\begin{center}
		{\includegraphics[scale=0.57,trim = 0.1in 0.75in 0in 0.4in, clip]{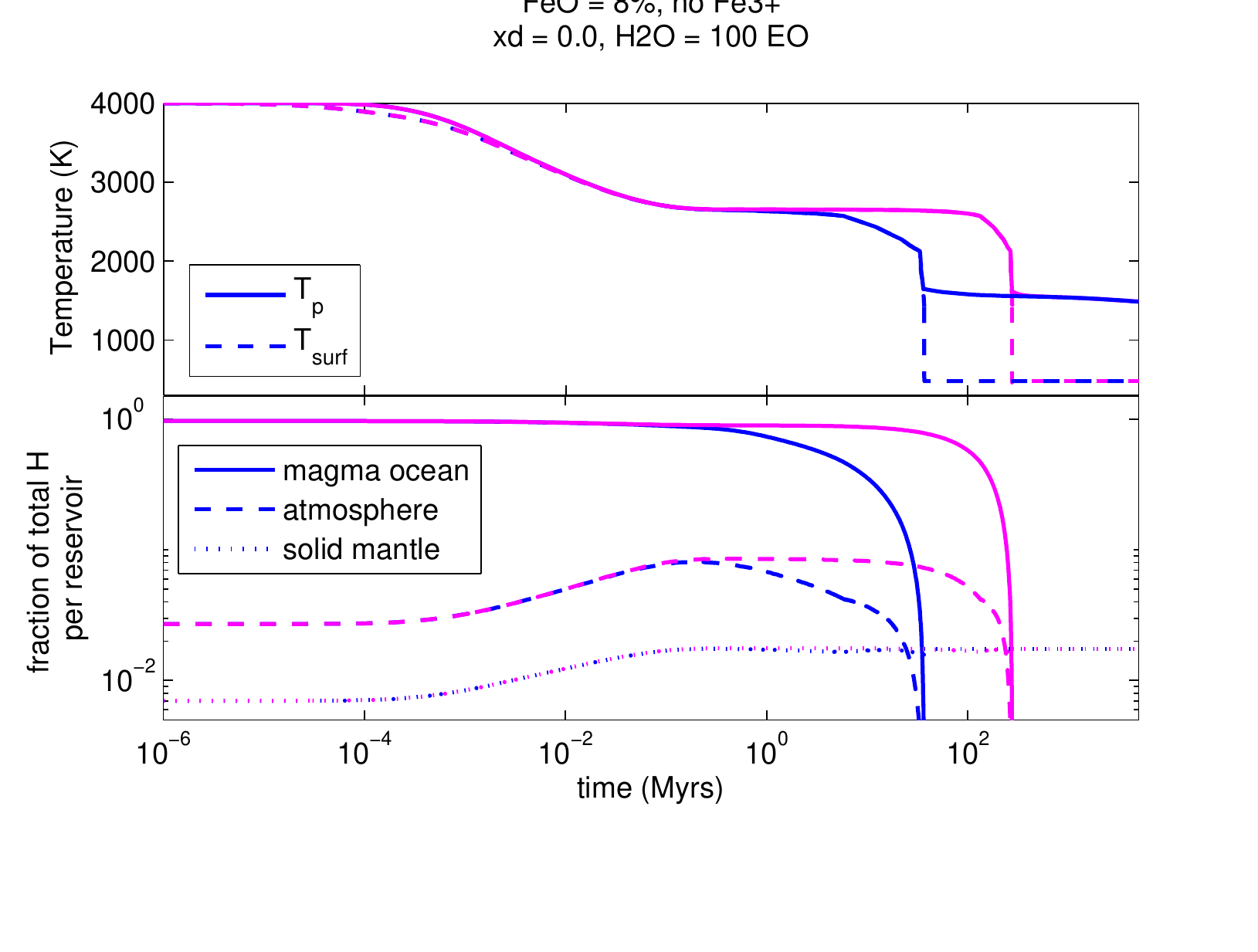}}
	\end{center}
	\caption{Sample model run for both XUV model A (blue) and B (pink). The top panel shows the evolution of the mantle potential temperature (solid line) and the surface temperature (dashed line). The bottom panel shows the evolution of the planetary water reservoirs (solid line: magma ocean, dashed line: atmosphere, dotted line: solid mantle). Model shown has FeO = 8 wt\%, $\chi_d$ = 0, and inital \ce{H2O} inventory of 100 Earth oceans (EOs).}
	\label{fig:modelrun}
\end{figure}

A sample run of the thermal model is shown in the top panel of Figure \ref{fig:modelrun}. The potential temperature and surface temperature are nearly identical until the "mush" stage is reached and the boundary layer begins to grow. When the surface temperature reaches the solidus temperature (1420 K), the magma ocean phase has concluded. When applied to an Earth-like planet, our thermal model reproduces the cooling times and heat fluxes found in \citet{Lebrun13} and \citet{Hamano13} very well. We deviate at later stages due to the fact that we do not include condensible atmospheric water vapor, which will not be present on GJ 1132b. However, the comparison gives us confidence that our thermal model produces reasonable results. 

\subsection{Volatile Model} \label{sec:volatile} 
Water is very soluble in silicate melts, so the \ce{H2O} pressure at the surface of the planet during the magma ocean stage is set by its solubility in the magma ocean. We use a fit to the solubility data of \citet{Papale97}:
\begin{equation}
p(Pa) = \left(\frac{F_{\ce{H2O}}}{3.44 \times 10^{-8}}\right)^{1/0.74}
\end{equation}
where $F_{\ce{H2O}}$ is the mass fraction of water in the liquid silicate melt. Note that the solubility of water within silicates at low pressures is effectively temperature independent. Mass balance for water within the magma ocean system is given by:
\begin{multline}
M^{mo,t}_{\ce{H2O}} = M^{crystal}_{\ce{H2O}} + M^{liq}_{\ce{H2O}} + M^{atm}_{\ce{H2O}} \\
= k_{\ce{H2O}}F_{\ce{H2O}}M^{crystal} + F_{\ce{H2O}}M^{liq} \\
+ \frac{4 \pi R^2_p}{g}(\frac{F_{\ce{H2O}}}{3.44 \times 10^{-8}})^{1/0.74}
\end{multline}
where $k_{\ce{H2O}}$ is the partition coefficient for water between melt and solid (0.01), $M^{mo,t}_{\ce{H2O}}$ is the mass of water in the magma ocean + atmosphere system on the current time step, and the mass of crystals ($M^{crystal}$) within the magma ocean is found from the melt fraction $\psi$ calculated along the adiabatic profile in the magma ocean. The total mass of the magma ocean ($M^{liq} + M^{crystal}$)is determined by difference with the radius of solidification. The mass of water in the magma ocean + atmosphere system ($M^{mo,t}_{\ce{H2O}}$) and the mass of water in the solid mantle at a given time are determined with the differential equations:
\begin{eqnarray}
\frac{dM^{sol}_{\ce{H2O}}}{dt} = 4 \pi \rho_m k_{\ce{H2O}} F_{\ce{H2O}} r^2_s \frac{dr_s}{dt}  \\
\frac{dM^{mo}_{\ce{H2O}}}{dt} = -\frac{dM^{sol}_{\ce{H2O}}}{dt} - 4 \pi r^2_p \phi_{1} \frac{\mu_{\ce{H2O}}}{2 \mu_{H}}
\end{eqnarray}
where $\phi_{1}$ is the XUV-driven atmospheric mass loss rate of H (in kg m$^{-2}$ s$^{-1}$), discussed in Section \ref{sec:primordial}. We find that our calculations conserve \ce{H2O} mass when atmospheric loss is turned off. A sample model run is shown in the bottom panel of Fig. \ref{fig:modelrun}. The figure shows the distribution of water between the three planetary reservoirs: magma ocean, atmosphere and solid mantle. When the magma ocean has cooled, most of the water remains in the solid mantle. Atmospheric escape is included in the calculation shown.

We assume that \ce{O2} is produced in the atmosphere by loss of H from \ce{H2O}. \ce{O2} is then also lost from the atmosphere at a slightly slower rate due to hydrodynamic drag, as discussed in Section \ref{sec:drag}. The O$_{2}$ produced in the atmosphere is also in contact with the FeO in the silicate melt. We allow the magma ocean to take up \ce{O2} by oxidation of FeO to \ce{FeO_{1.5}}. The equilibrium oxygen fugacity for the magma ocean is given by \citet{Kress91}:
\begin{multline}
ln\left(\frac{X_{\ce{Fe2O3}}}{X_{FeO}}\right) =  0.196ln\left(f_{\ce{O2}}(Pa)\right) + \frac{11,492}{T} - 6.675 \\
-2.243X_{\ce{Al2O3}} -1.828X_{FeO^*} + 3.201X_{CaO} + 5.854X_{\ce{Na2O}}\\
+ 6.215X_{\ce{K2O}} - 3.36\left[1-\frac{1673}{T}-ln(\frac{T}{1673})\right] \\
- 7.01 \times 10^{-7}\frac{p(Pa)}{T} - 1.54 \times 10^{-10} \frac{(T-1673)p(Pa)}{T}  \\
+ 3.85 \times 10^{-17} \frac{p^2(Pa)}{T}
\end{multline}
where $X_{i}$ is the molar concentration of the oxides in the silicate melt, and we use the Bulk Silicate Earth (BSE) as our nominal composition \citep{ONeill1998}. This empirical relationship was derived for a wide range of natural silicate melt compositions equilibrated at oxygen fugacities ranging from metal-silicate equilibrium (iron-wustite buffer) up to air at temperatures between 1473 and 1900 K. Rocky exoplanets are expected to have relatively similar rocky elemental abundances to the Earth based on analysis of the observed mass-radius measurements and stellar elemental abundances \citep[][and references therein]{Dressing2015}. GJ 1132 b itself falls very close to an Earth-composition track on the mass-radius diagram \citep{Berta15}. The empirical oxygen fugacity relationship therefore is likely to cover the relevant compositional range for rocky exoplanets. The strongest mantle influence on the oxygen fugacity will likely be the total abundance of FeO (FeO*) in the silicate, so we explore a range of FeO abundances in our calculations. Note that we assume a metal-free magma ocean for these calculations. 

Mass balance between the atmosphere and the magma ocean is calculated for \ce{O2} in the same way as for \ce{H2O}. Oxygen is sequestered into the solid mantle as \ce{FeO_{1.5}}. We assume no fractionation between liquid and solid, either for FeO or \ce{FeO_{1.5}}, although Fe is known to fractionate from Mg in the melt (i.e., minerals that condense early should be less Fe-rich than those that condense later). We consider that this will have only a small effect on our oxygen mass balance, but we discuss implications in a later section. The magma ocean + atmosphere system loses oxygen to the solid mantle and to atmospheric escape, while atmospheric escape of H from \ce{H2O} produces O. This gives us two more differential equations for the abundance of free O in the solid and in the magma ocean + atmosphere system:
\begin{eqnarray}
\frac{dM^{solid}_{O}}{dt} = 4 \pi \rho_m F_{\ce{FeO_{1.5}}} r^2_s \frac{dr_s}{dt} \frac{\mu_O}{2 \mu_{\ce{FeO_{1.5}}}} \\
\frac{dM^{mo}_{O}}{dt} = 4 \pi R^2_p \phi_{1} \frac{\mu_O}{2 \mu_H} - 4 \pi R^2_p \phi_2 - \frac{dM^{solid}_{O}}{dt}
\end{eqnarray}
where $\phi_2$ is the XUV-driven atmospheric mass loss rate of O (in kg m$^{-2}$ s$^{-1}$) and $F_{FeO_{1.5}}$ is the mass fraction of \ce{FeO_{1.5}} in the mantle. Following magma ocean solidification, direct exchange of oxygen between the mantle and the atmosphere halts ($dM^{solid}_{O}/dt = 0$). Oxygen no longer exchanges with the mantle following solidification, but is continuously created by H loss and lost by hydrodynamic drag. Although similar models for Venus have allowed continued O loss due to oxidation of the crustal layer \citep{Gillmann2009}, we consider this effect to be small given that the upper mantle will already be significantly oxidized by exchange with the atmosphere during the magma ocean phase. Water outgassing continues but there is no return of water to the mantle after magma ocean solidification. Outgassing is parameterized similar to \citet{Sandu2012}:
\begin{equation}
r_{outgas} = 4 \pi r_{p}^{2} \rho_{m} F_{melt}^{avg} f_{melt}^{avg} u_{c} \chi_{d}
\end{equation}
where $F_{melt}^{avg}$ is the volume-averaged mass fraction of water in the melt, $f_{melt}^{avg}$ is the volume-averaged melt fraction of the mantle, $u_{c}$ is the mantle convection velocity, and $\chi_{d}$ is the degassing efficiency, which can vary from 0 (no degassing) to 1 (completely efficient degassing). We will explore the effect of the degassing efficiency in our discussion of the results.

\subsection{Properties of GJ 1132b} \label{sec:planet}

The mass and radius for GJ 1132b are taken from the discovery paper \citet{Berta15}. The core mass and radius assuming a two-component model (silicate + metal, no water) are determined with the online tool of \citet{Zeng16}. Values for the planet properties are given in Table 1. The mass of the planet is currently only known to 3$\sigma$, although continued Doppler monitoring will shrink the mass uncertainty and enable more detailed compositional models. We note that there is additionally a well-known degeneracy in determining the planet's composition from the density. Using the online tool of \citet{Zeng16}, we find that the nominal mass of the planet allows for up to about 20 wt\% of the planet to be water. Note that this extreme value results in a nearly zero silicate mass fraction, which is highly unphysical, as giant impact simulations show that mantle stripping can produce planets with at most 70\% core mass fraction \citep{Marcus2010}. However, we will test loss models here for total planetary water abundances up to 20 wt\% as a limiting case, while holding the core and silicate mass fractions fixed at the value determined assuming the present measured mass and a two-component silicate-metal model. 
\begin{table}
	\centering
	\caption{Parameters for GJ1132b used in the modeling.}
	\begin{tabular}{ll}
		\hline
		\hline
		Parameter & Values \\
		\hline
		Present-day stellar luminosity $L$ [$L_\odot$] & 0.00438 \\
		Orbital distance $a$ [AU] & 0.0153 \\
		Planetary mass $M_p$ [$M_\oplus$] & 1.62 \\
		Planetary radius $r_p$ [$r_\oplus$]  & 1.16 \\
		Core mass fraction $M_c$ [$M_p$] & 0.262 \\
		Core radius $r_c$ [$r_p$] & 0.54 \\
		Surface gravity $g_0$ [m~s$^{-2}$] & 11.8 \\
		Planetary albedo $A$ & 0.75 \\
		\hline
		\hline
	\end{tabular}\label{tab:params}
\end{table}

\section{Results of Coupled Models} \label{sec:results}
We explore model results for the two XUV flux models. For both models, we vary the initial planetary water inventory and the mantle FeO abundance. We explore water inventories ranging from 0.1 up to 1000 Earth oceans (EO = $1.39 \times 10^{21} kg$) of water, which is about 20 ppm in the mantle up to about 20 wt\%. We note that while there are measurements of the solubility of water in silicate melts up to this value, the data beyond 10 wt\% is sparse and fairly poorly constrained. 

For mantle FeO, we consider abundances ranging from 0.1 to 20 wt\%. Abundances of FeO in the silicate mantles of the terrestrial planets in the Solar System span this range. Estimates for Mercury's mantle are 2-3 wt\%, Earth and Venus have about 7 - 8 wt\% of FeO, whereas Mars has a mantle FeO abundance of about 18 wt\%, and Vesta 20 wt\% \citep{RobinsonTaylor}. The abundance of FeO in the mantle is a result of the composition of the protoplanets out of which a planet is made, the conditions under which core formation occurs and any subsequent reducing or oxidizing processes. The abundance of \ce{Fe^3+} (or \ce{FeO_{1.5}}) in the Earth's mantle is fairly small (\ce{Fe^3+}/Fe$^{total} \sim 0.02 - 0.03$, \citet{FrostMcCammon}). We examine values from 0 up to 0.03, and find only a minor difference on the final results of the model.  Our nominal results use \ce{Fe^3+}/Fe$^{total}$ = 0, which gives the mantle maximum oxygen uptake potential. 

We also consider the effect of efficient ($\chi_{d} = 1$) versus inefficient ($\chi_d = 0$) degassing after the magma ocean stage has solidified. This parameter has an effect on the final water and \ce{O2} abundances, as we discuss below.

\subsection{Water loss and Magma Ocean Solidification} \label{sec:tmo}
\begin{figure}
		\begin{center}
			{\includegraphics[scale=0.6,trim = 1.5in 3in 0in 3in, clip]{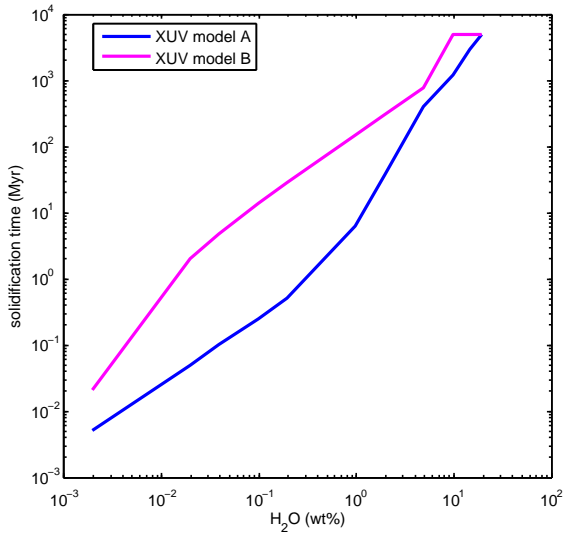}}
		\end{center}
		\caption{Solidification time (in Myr) for different initial water abundances. Water abundances are weight percent of the total planet. The figure shows results for our two XUV flux models, A (blue) and B(pink). For lower XUV fluxes (model B), the magma oceans are longer-lived. For water abundances greater than ~10 wt\%, the magma oceans persist for the entire length of our calculation (5 Gyr). }
		\label{fig:solidification}
	\end{figure}

	Magma ocean solidification times depend strongly on the initial water abundance of the planet, as well as the XUV flux. Figure \ref{fig:solidification} shows the solidification times for the two XUV models as a function of initial water abundance. Models were run for a total integration time of 5 Gyr, consistent with the estimated age of GJ 1132 \citep{Newton2016}. The XUV flux model B results in magma oceans that persist roughly an order of magnitude longer than for XUV flux model A. The longer duration is due to the slower loss of water vapor from the atmosphere, which causes the planet to remain hotter for longer.
	
\begin{figure}
		\begin{center}
			{\includegraphics[scale=0.6]{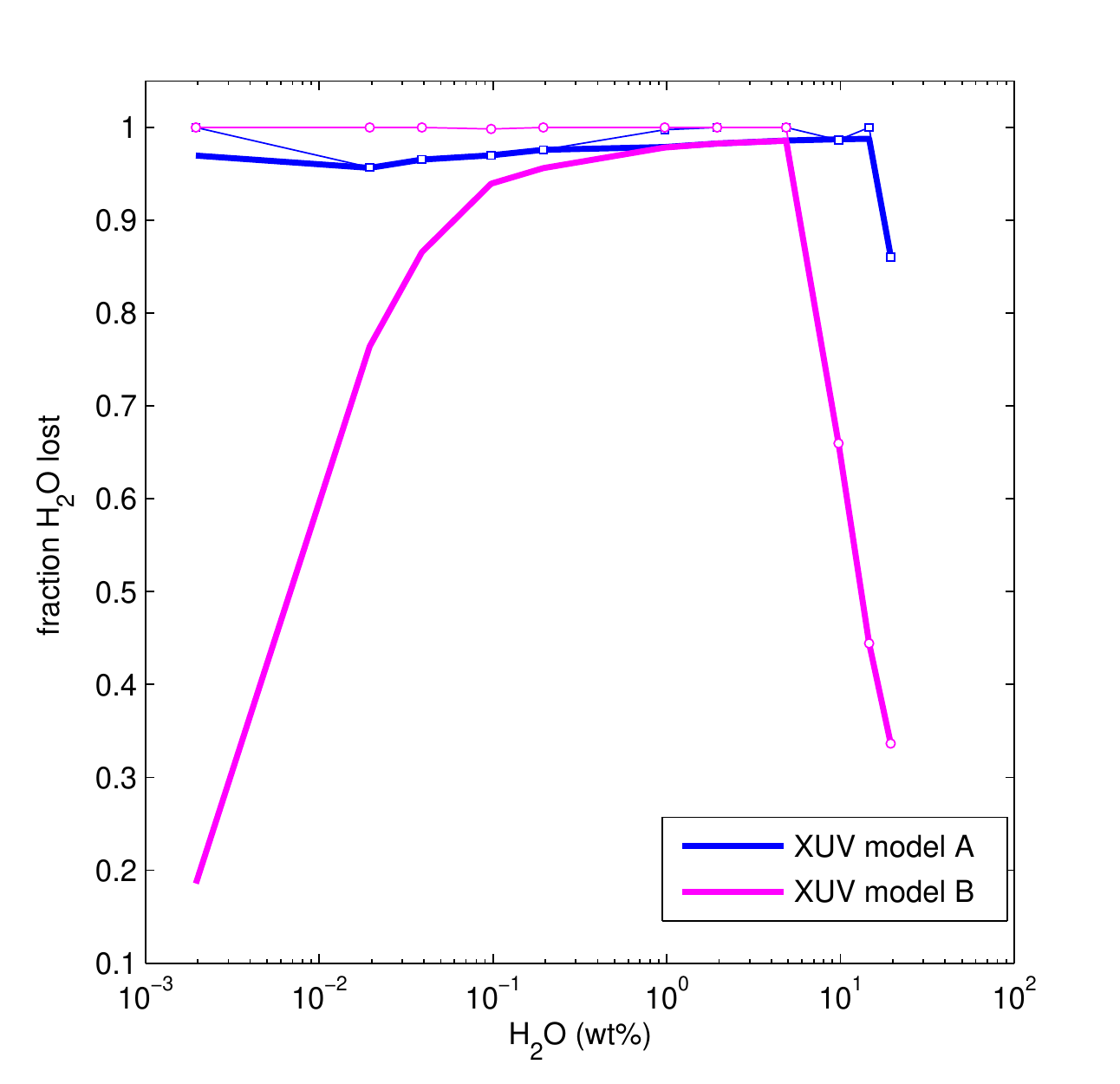}}
		\end{center}
		\caption{Fraction of total water lost as a function of initial planetary water abundance. The figure shows results for the two XUV flux models, A (blue) and B(pink). Thick lines are for $\chi_{d} = 0$ (i.e. no outgassing after magma ocean solidification), whereas thin lines with open points (A:square, B:circle) are for $\chi_{d} = 1$ (i.e. perfectly efficient outgassing after magma ocean soldification). The difference between the thin and thick lines for XUV model B indicates that most of the planet's water is lost after magma ocean soldification. }
		\label{fig:H2Olost}
	\end{figure}

	Total planetary water loss also depends strongly on both the initial water abundance and the XUV flux, as well as the degassing efficiency in the post-magma ocean state. Figure \ref{fig:H2Olost} shows the fraction of intial water lost for both flux models, as well as different degassing efficiencies. For XUV flux model A, water loss is more than 95\% complete for all intial water abundances, except the largest. For XUV flux model B, the figure shows that the amount of water lost for low initial water abundances depends strongly on the degassing efficiency. For efficient degassing, all water is lost except for initial water abundances $\geq$ 10 wt\% of the planet. For no degassing post-magma ocean, the fraction of the initial water lost increases with increasing water abundance, but only up to 10wt\%, above which water loss decreases. This indicates that for low water abundances, most of the remaining water is stored in the mantle and is lost after the magma ocean phase during passive outgassing of the interior. If degassing is inefficient, the water can be permanently trapped in the mantle. A large melting event, possibly caused by late impacts, could induce further outgassing. The amount of remaining water is not sufficient to affect the planet's density except at the very highest water abundances where most of it remains in the atmosphere. For XUV model B and no degassing, the remaining water abundance is $\sim$10 wt\% of the planet's mass. This is technically consistent with the present mass and radius measurement, but requires un-Earth-like silicate(0.36) and core(0.54) mass fractions \citep{Zeng16}. The majority of the water in this scenario would be locked in solid phases in the mantle. 
	
\subsection{Atmospheric oxygen}
		\begin{figure*}
			\begin{center}
				\includegraphics[scale=0.65,trim = .8in 0in 0in 0in, clip]{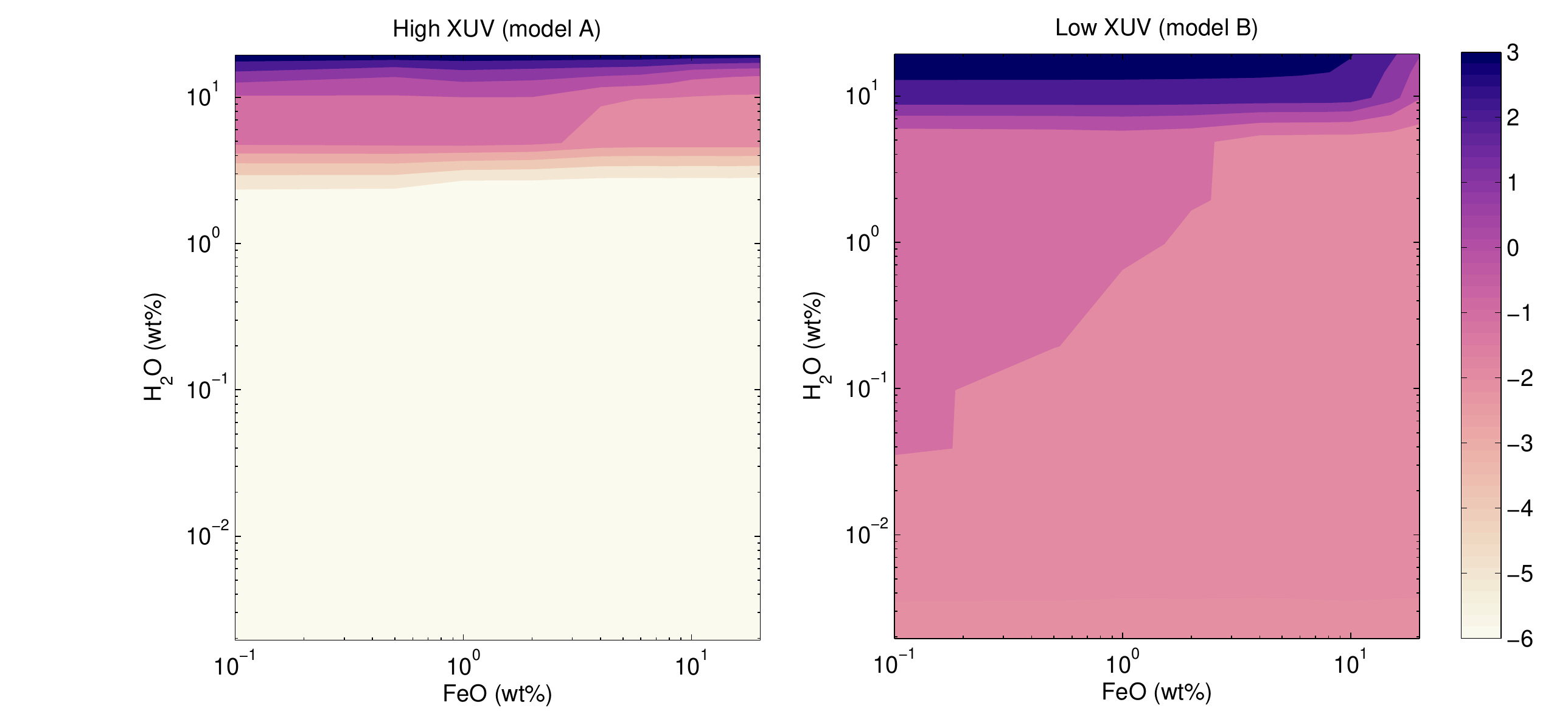}
			\end{center}
			\caption{Final \ce{O2} pressure in the atmosphere (in $log_{10}$(bars)) as a function of initial \ce{H2O} and FeO contents for both XUV flux models (left: high XUV, right: low XUV). Here we set $\chi_d = 0$.}
			\label{fig:finalO2_1}
		\end{figure*}
	
	For models with intial $Fe^{3+}/Fe^{total} = 0$, all free oxygen is produced by destruction of water and loss of hydrogen. The amount of oxygen produced is therefore directly proportional to the amount of water lost, shown in Fig. \ref{fig:H2Olost}. XUV model A therefore typically produces more total oxygen than model B, especially at high initial water abundances and at lower abundances when degassing is inefficient.
	 
\begin{figure}
			\begin{center}
				\includegraphics[scale=0.65,trim = 5.25in 0in 0in 0in,clip]{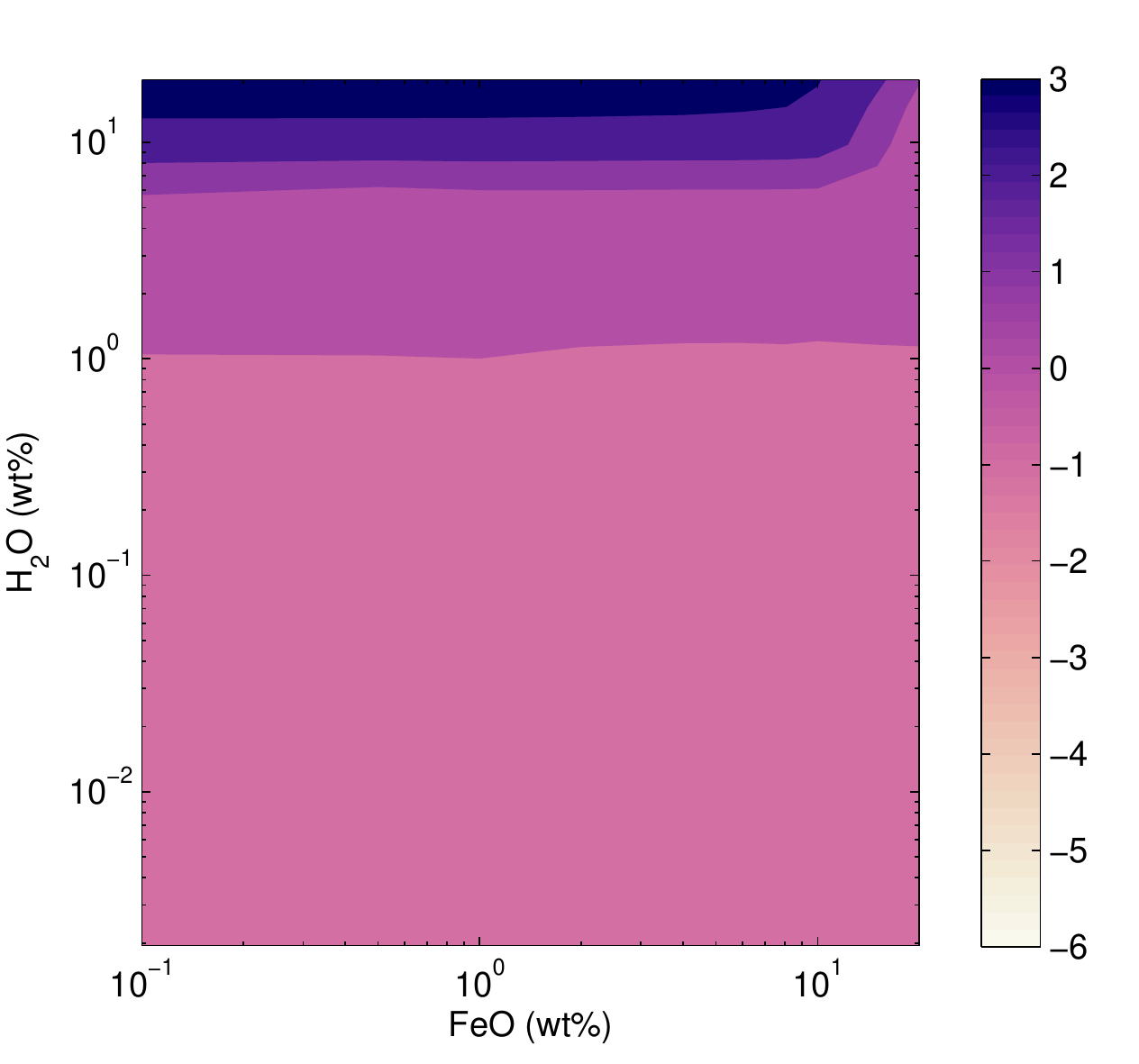}
			\end{center}	
			\caption{Same as Figure \ref{fig:finalO2_1} for XUV model B with efficient degassing ($\chi_{d} = 1$). Degassing efficiency makes no difference on the final atmospheric \ce{O2} abundance for XUV model A (not shown). }
			\label{fig:finalO2_2}
		\end{figure}

	A fraction of the oxygen that is produced is lost both to atmospheric escape and to the mantle. For both XUV models, between 90 to 99 \% of the total oxygen produced by photolysis of water vapor is lost to space, with higher O loss amounts occurring at low water abundances. At most 10\% of the total oxygen produced is sequestered into the mantle (discussed below). The remainder of the oxygen resides in the atmosphere, as shown in Figure \ref{fig:finalO2_1}. Both XUV models can result in residual \ce{O2} remaining in the atmosphere at a few bar level, with minimal dependence on FeO content. Several hundred to several thousand bars can build up only for initial water abundances greater than about 5 wt\% of initial water. For XUV model A, the final atmospheric \ce{O2} abundance is negligible for water abundances below $\sim$5-10 wt\% initial water, whereas for XUV model B, the \ce{O2} abundance in the atmosphere is slightly dependent on the FeO content of the mantle, with more \ce{O2} atmospheric build up for smaller FeO abundances. This is because there is a smaller sink for \ce{O2} in the magma ocean with lower FeO abundances. Degassing efficiency affects the final \ce{O2} abundance in the atmosphere for XUV model B with \ce{H2O} abundances less than 10 wt\% as shown in Figure \ref{fig:finalO2_2}. At higher water abundances, persistent magma oceans mean that the model never enters the passive degassing state. For water abundances less than 10 wt\%, the atmosphere has about 10 times more oxygen than for inefficient degassing. This is due to creation of additional oxygen by dissociation of water degassed in the post-magma ocean time frame. 

	We find that for both XUV models, \ce{O2} is more abundant than water vapor in the atmosphere  for nearly all of our parameter space, but the atmosphere is likely to be fairly tenuous ($p < few ~bar$). Steam dominates the atmosphere only for XUV model B at the highest water abundances, with about a factor of 10 more water vapor than \ce{O2}. Therefore our models indicate that the atmosphere of GJ 1132b may be tenuous and dominated by \ce{O2}. If abundant atmospheric water is observed, it is indicative of both a low XUV flux history and high initial abundance. 
	
	For mantles with initial \ce{Fe^{3+}}/Fe$^{total}$ of 0.02 - 0.03, we find that atmospheric \ce{O2} is relatively unaffected. At low water abundances, \ce{O2} is the same as for our nominal calculations. At large water abundances ($> 5$ wt\% for XUV B), atmospheric \ce{O2} is the same for low initial FeO, but is slightly larger than for our nominal model as FeO increases. We find a maximum increase at 20\% initial FeO of about 50\% in the \ce{O2} atmospheric pressure. 
	
	Planetary albedo has a slightly larger effect on our results. For a lower planetary albedo of 0.3, we find for XUV model B that the final \ce{O2} atmospheric pressure is 60 - 90\% of our nominal results. For XUV model A, results are the same (i.e., p $<<$ 1)at water abundances below  5 wt\%, and are about 75 - 90\% of the nominal results at higher water abundances. In both cases we find that the fraction of the nominal abundance increases with increasing water abundance. That is, albedo has a larger effect on models with lower initial water abundances. However, the effect of albedo is small enough that it does not alter our primary conclusions.  
	 		
\subsection{Mantle Composition}
	\begin{figure*}
		\begin{center}
			\includegraphics[scale=0.65,trim = 0.5in 0in 0.0in 0in, clip]{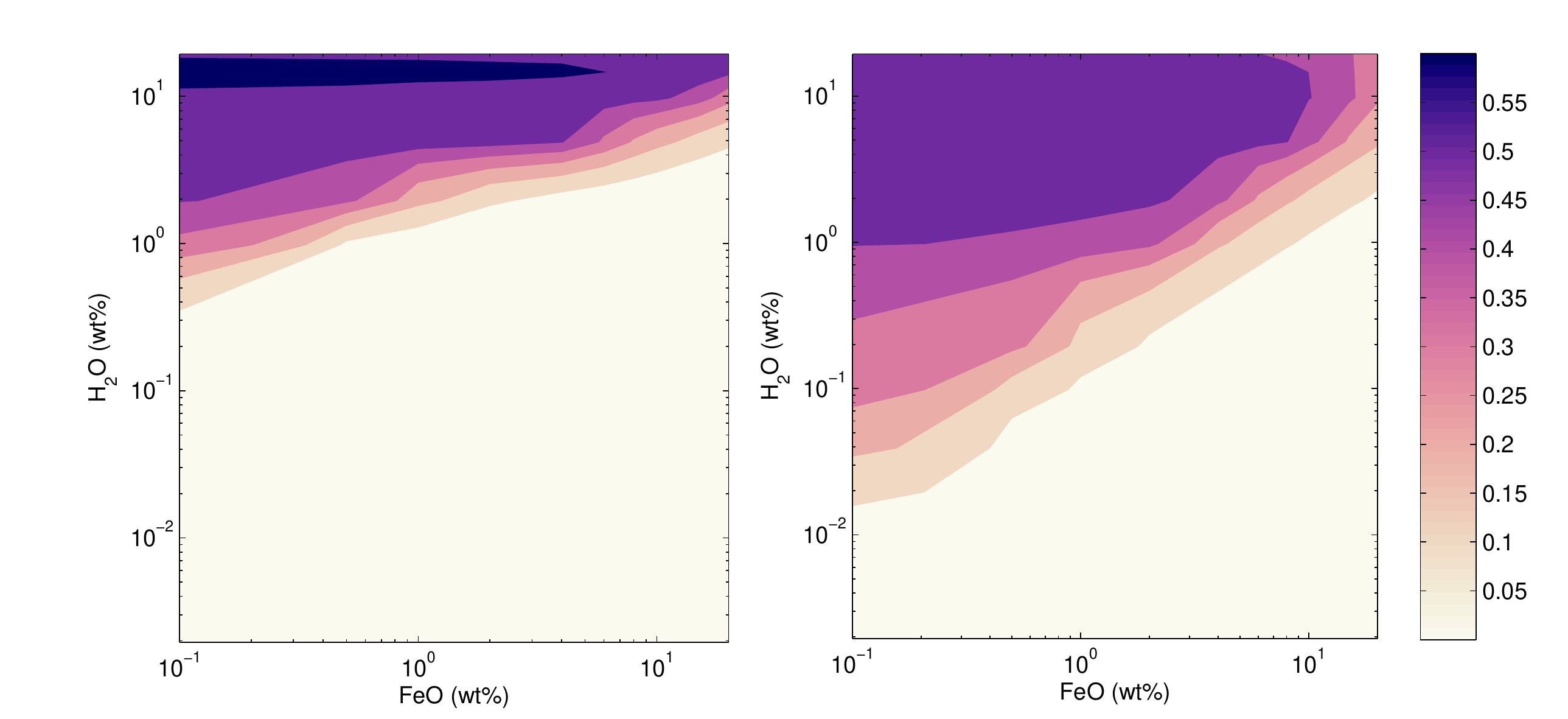}
		\end{center}
		\caption{Mantle averaged ratio of \ce{FeO_{1.5}} to initial FeO in the mantle as a function of initial \ce{H2O} and FeO contents for both XUV flux models (left: XUV A, right: XUV B). We include \ce{FeO_{1.5}} remaining in the magma ocean for those models which do not fully solidify.}
		\label{fig:Fe2O3}
	\end{figure*}

While more extensive destruction of \ce{H2O} for XUV model A implies greater production of oxygen, most of the oxygen is directly lost to space and therefore a smaller fraction  can be absorbed into the mantle than for XUV model B. For XUV model A, this is at most 8\% of the total oxygen produced, whereas slightly more (10\%) can be absorbed for model B. For XUV model A extensive absorption of O by the mantle only occurs at high water ($>$ 5 wt\%)and FeO abundances ($>$5 wt\%). In contrast, extensive oxygen absorption occurs across wide ranges of water ($>$0.05 wt\%) and FeO ($>$ 0.5 wt\%) for XUV model B, although the highest relative absorption still occurs at the largest FeO and water abundances. 
	
While high FeO abundances lead to more oxygen absorption, the conversion (or oxidation) of FeO to \ce{FeO_{1.5}} is more extensive at low FeO abundances as shown in Figure \ref{fig:Fe2O3}. Note that we include here the \ce{FeO_{1.5}} remaining in the magma ocean for those models which do not fully solidify. For XUV model A, the peak oxidation occurs above 15 wt\% \ce{H2O}, at FeO abundances less than 8 wt\%. For XUV model B, the peak is at 5 wt\% \ce{H2O} for FeO abundances less than about 5 wt\%. Less oxidation occurs for XUV model B at higher \ce{H2O} abundances because these models lose less H from the atmosphere and therefore produce less free O.  

\begin{figure}
		\begin{center}
			\includegraphics[scale=0.65]{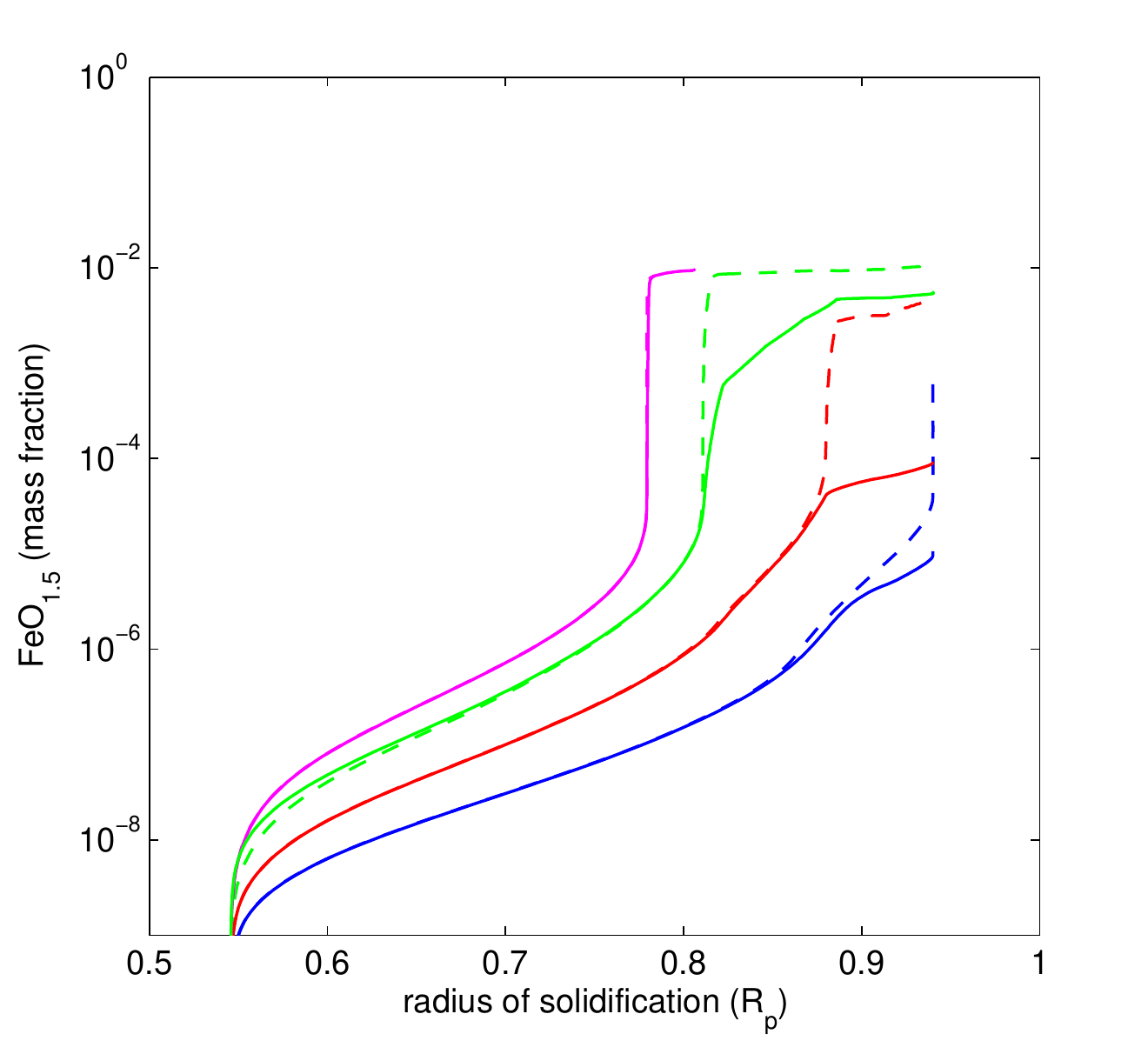}
		\end{center}
		\caption{Abundances of \ce{FeO_{1.5}} in the mantle with depth in the solidifying magma ocean starting at the core mantle boundary. Solid lines are for XUV model A, dashed lines for XUV model B. Colors refer to the planetary water abundance (blue: 1 EO, red: 10 EO, green: 100 EO, pink: 1000 EO). For the models shown, the total FeO abundance in the mantle is 1 wt\% with no initial \ce{Fe2O3}. Results are shown at the end of the integration time of 5 Gyr, so some magma oceans are not fully solidified. Additionally, magma ocean solidification stops when the surface temperature reaches the solidus, but there is still a substantial melt layer from 0.94 $R_{p}$ to the base of the thermal boundary layer. This is why none of the curves extend to a full planetary radius. For non-zero initial abundances, less stratification in mantle composition is observed, with most of the lower mantle solidifying with the initial \ce{FeO_{1.5}} abundance.}
		\label{fig:Fe2O3profile}
	\end{figure}

Figure \ref{fig:Fe2O3profile} shows the profile of \ce{FeO_{1.5}} abundance with depth in the solidifying magma ocean at the end of the integration period of 5 Gyr. The remaining liquid at larger radii has the same \ce{FeO_{1.5}} abundance as the last layer of solidified mantle. The outer radius of the magma ocean is smaller than the planetary radius because of the formation of a thermal boundary layer, which insulates the upper mantle. The maximum abundance of \ce{FeO_{1.5}} is limited by the total FeO content, which is fixed at 1 wt\% in this figure. As can be seen, the mantle becomes progressively more oxidized as the magma ocean solidifies, and the degree of oxidation is strongly dependent on the total water abundance. Less stratification occurs for non-zero \ce{Fe^3+}/Fe$^{total}$ starting abundances. The progressive oxidation of the mantle may effect later mantle convection. The density of silicates enriched in \ce{FeO_{1.5}} will be slightly lower than more reduced silicates, which results in a stably stratified mantle. This may delay the onset of solid state convection after magma ocean solidification. We will discuss this possibility in the next section. 

\section{Discussion} \label{sec:discussion}

\subsection{Sensitivity of loss rate to atmospheric composition}\label{sec:sensitivity}
\begin{figure}
	\begin{center}
		\includegraphics[scale=0.65]{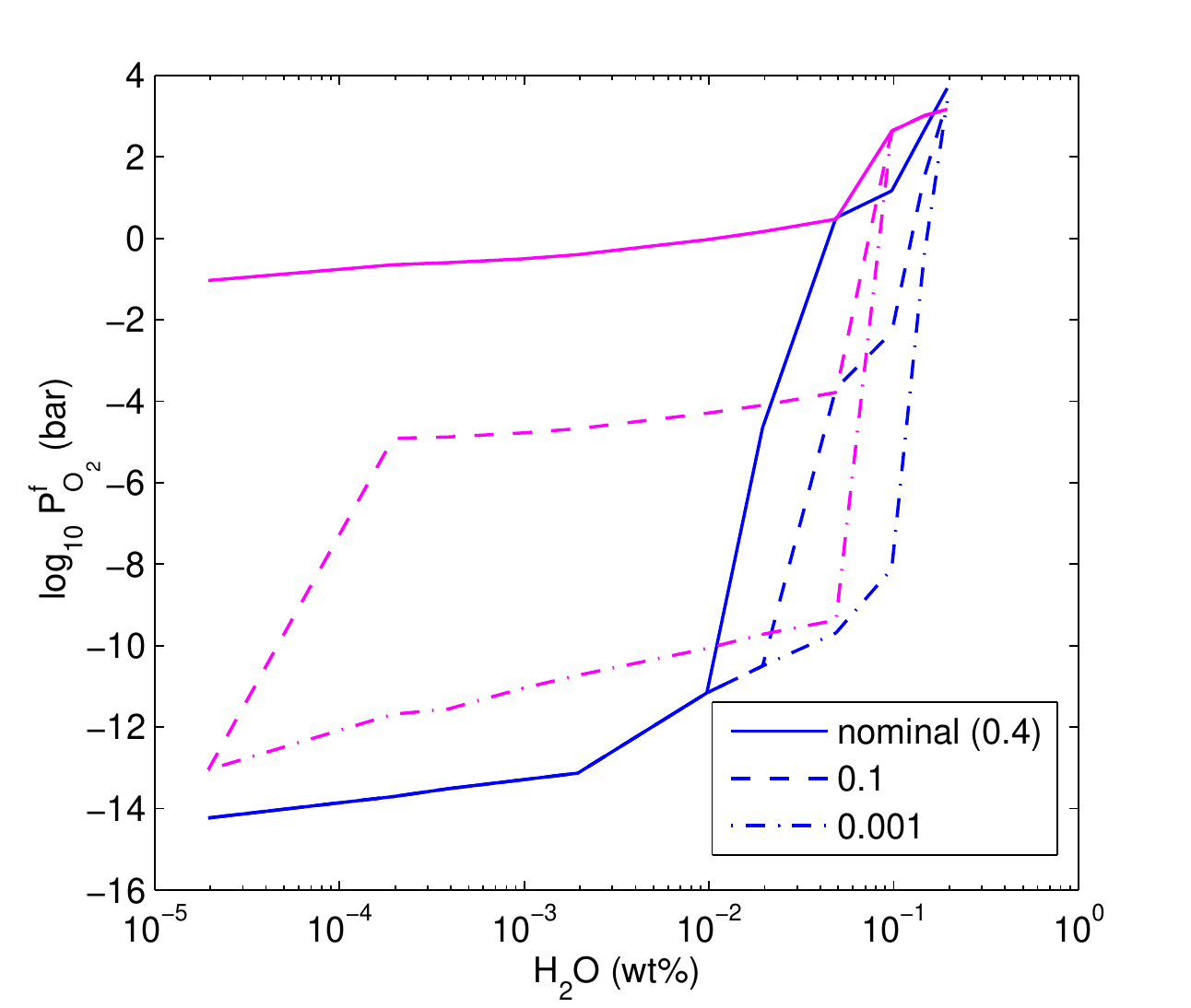}
	\end{center}
	\caption{Sensitivity of the final \ce{O2} pressure to the transition point between energy-limited loss of H (with hydrodynamic drag of O) to diffusive-loss of H through an O background gas (with no loss of O). The transition composition is given in terms of the molar abundance of H in the atmosphere, calculated from the total pressures of \ce{O2} and \ce{H2O} ($X_H = 0.4(nominal), 0.1, 0.001$). Blue lines are for XUV model A, pink lines are XUV model B. }
	\label{fig:crossover}
\end{figure}

In our nominal models, we assume that atmospheric loss is energy-limited, where the loss rates are dependent on the O and H molar concentrations. We assume that energy-limited escape driven by hydrodynamic loss of H occurs until the \ce{O2} and \ce{H2O} total atmospheric pressures are equal. After this cross-over point, we assume that H must diffuse through the O background gas, at which point the hydrodynamic loss halts and O no longer escapes. However, the transition composition is uncertain because H should diffuse more readily into the upper atmosphere than O. We explore the sensitivity of our final results to this transition point in Figure \ref{fig:crossover}. Here we show results for both XUV flux models with $\chi_d = 1$ for a constant FeO abundance of 8 wt\% as a function of initial water abundance for different transition points ($X_H$ = 0.4(nominal), 0.1, and 0.001). For XUV model A, the final \ce{O2} pressure is insensitive to the transition point up to $\sim$1 wt\% of \ce{H2O}. At higher water abundances, the final \ce{O2} pressure is reduced by several orders of magnitude as the transition point drops, except at the very highest water abundance where the magma ocean persists. For XUV model B, the transition point has a strong effect on the \ce{O2} abundance for initial water abundances less than $\sim$10 wt\%. Reducing the transition abundance results in more tenuous \ce{O2} atmospheres, since more of the O can escape. 

\subsection{Loss of an earlier \ce{H2} envelope}
It is possible that GJ 1132b began with an envelope dominated by \ce{H2} gas, rather than \ce{H2O}. As discussed in Section \ref{sec:primordial}, a significant mass of \ce{H2} can be lost from the planet, up to 15\% of the planet's mass over 10 Gyr. Interaction of an \ce{H2} atmosphere with mantle FeO might result in reduction of mantle FeO to Fe metal through a reaction such as:
\begin{center}
	\ce{H2(g) + FeO <=> H2O(g) + Fe}
\end{center}
The forward reaction is thermodynamically unfavorable and has been shown to go nearly to completion in the reverse direction (oxidation of metal) at all temperatures and pressures on the present and early Earth \citep{Fukai84,Kuramoto1996}. In fact, these experimental studies of the iron-water reaction at high pressure have shown that hydrogen liberated from water can be sequestered into a \ce{FeH_x} metallic phase via a reaction such as:
\begin{center}
	\ce{H2O(g) + 2Fe -> FeO + FeH_x}.
\end{center}
However we expect that relative to the total duration of the magma oceans, the presence of metal within the magma ocean was relatively short-lived. We therefore consider that the primary effect of an initial \ce{H2}-envelope would be to prolong the magma ocean lifetimes and reduce the loss of water and \ce{O2} from those calculated here. 

\subsection{Effect of \ce{CO2}} \label{sec:CO2}
\ce{CO2} is a common atmospheric component that is often included in magma ocean models \citep{ET08,Lebrun13}, due to both its large abundance and its contribution to greenhouse warming. We do not consider it here in order to minimize the complexity of the model, but we will qualitatively discuss its possible effect on the evolution of GJ 1132b. The solubility of \ce{CO2} in silicate melt is much lower than that of \ce{H2O}, but it is much more soluble in metal alloy. Therefore, numerous papers on Earth-based magma ocean models have noted that \ce{CO2} will be concentrated in either the atmosphere or the core. \citet{Hirschmann2012} argues based on alloy/melt partition coefficients that a magma ocean that equilibrates with only 1 wt\% of alloy would lose at least 60\% of its total carbon to the core. However, we noted above that the presence of metal within the magma ocean was likely relatively short-lived, so unless carbon is removed during core formation, it seems likely that there should be substantial carbon remaining in the magma ocean and atmosphere. 

Solubility of \ce{CO2} in the mantle depends on the temperature, pressure, melt composition, and oxygen fugacity: as GJ 1132 b becomes more oxidized, \ce{CO2} should become more soluble in the melt. However, solubility relationships indicate that it is unlikely that more than about 20 - 30\% of the \ce{CO2} could be dissolved in the magma ocean \citep{Holloway1998}. \citet{Hirschmann2012} also suggests the possibility of diamond precipitation in the mid to lower mantle or a magma ocean carbon pump to lower atmospheric \ce{CO2} abundances. This would sequester carbon in the mantle where it would be available for later outgassing during a post-magma ocean state, much like water in our efficient degassing scenarios. However, while this is a possibility, it would require detailed additional modeling to evaluate. 

\ce{CO2} in the atmosphere will prolong the magma ocean lifetimes by additional greenhouse warming, which may enhance atmospheric loss of both water vapor and \ce{O2}. \citet{Tian2009} showed that in highly irradiated super-Earth atmospheres dissociation of \ce{CO2} can lead to both carbon and oxygen loss, with carbon escaping more rapidly due to its lower atomic weight. \citet{Wordsworth13} showed that water loss from \ce{CO2}-rich atmospheres can still be substantial, especially for planets that receive more insolation than the present day Earth, such as GJ 1132 b. While \ce{CO2} is effective at cooling the upper atmosphere, which can hinder loss in more temperate planets, a back-of-the-envelope calculation suggests that the degree of cooling from the \ce{CO_2} 15 $\mu$m non-LTE emission would still be far lower than the XUV flux received by GJ 1132 b, at least for the first Gyr. Therefore, cooling of the upper atmosphere would likely be insufficient to hinder the escape of \ce{O_2} and \ce{CO2}. Therefore, tenuous \ce{O2} atmospheres are the most likely scenarios for GJ 1132 b after loss of an \ce{H2} envelope. Non-thermal effects provide additional loss avenues as discussed below.

\subsection{Non-thermal loss mechanisms}
Considering non-thermal mechanisms for atmospheric escape from GJ1132b, it is safe to assume no planetary magnetic field, in analogy to Venus and as a conservative choice. Also, while GJ1132b is closer than Venus to its star in terms of bolometric irradiation and stellar wind flux, it is significantly more massive. As a result, charge exchange (e.g., $H + H^{+*} \rightarrow H^{+} + H^{*}$) and ion escape could increase the loss of hydrogen, and dissociative recombination (e.g., $O_{2}^{+} + e^{-} \rightarrow O^{*} + O^{*}$) could increase the loss of oxygen. The latter mechanism releases only $0.6 \times 10^{-18}$ J per atom, which is not enough to permit escape given the high mass of GJ1132b. The hydrogen non-thermal escape might be insignificant, by analogy to Venus (e.g. Pierrehumbert 2010), but a dedicated study is warranted. Similarly, the stellar wind of M dwarfs like GJ1132 is expected to be too tenuous to lead to significant atmospheric erosion, but no firm conclusion is possible without detailed modeling \citep{Kislyakova2013,Kulikov2006}.

\subsection{Mantle convection after soldification}
In order to have efficient degassing during the post-magma ocean state, the mantle of GJ 1132 b must continue to convect. However, progressive oxidation of the mantle by liberated O should lead to lower density materials at the top of the mantle, as shown in Figure \ref{fig:Fe2O3profile}. This may prevent overturn of mantle and delay the onset of solid state mantle convection, which would lead to reduced outgassing efficiencies. However, \citet{ET03} calculated the mineralogy of a solidifying magma ocean (without atmospheric oxidation), and found that the cumulate pile of the solidified magma ocean is unstable due to the partitioning of FeO into later (near surface) crystal phases. 

Although oxidation of FeO to \ce{FeO_{1.5}} may change the exact mineral condensation sequence, the additional oxygen should not be sufficient to counteract the density effect of enhanced FeO abundance in the upper mantle. For low FeO abundances, the density instability may be insufficient to cause mantle overturn, in which case GJ 1132 b may become stuck in a stagnant lid regime. This would mimic the low degassing efficiency model, which we have shown is only important in the case of the low XUV model B. Inefficient degassing reduces the final \ce{O2} abundance by about an order of magnitude in pressure.

\section{Predictions for GJ 1132b}\label{sec:predictions}
Our model suggests that GJ 1132 b would require more than $\sim$5 wt\% by planet mass of initial water in order to retain a substantial steam envelope. Substantial oxygen atmospheric abundances (a few bars up to several thousand bars) without significant steam ($<<$ 1 bar) would imply a relatively high XUV flux and initial water abundances greater than $\sim$5 wt\%. Substantial oxygen atmospheric abundances ($>$ 500 bars) with significant steam ($>$ 500 bars) would imply either low XUV flux over the system's lifetime, a large initial water abundance of more than 250 EO, or the presence of an earlier \ce{H2}-rich envelope. The presence of a steam atmosphere implies the continued existence of a magma ocean at GJ 1132 b's surface. However, most of our starting conditions result in tenuous atmospheres with at most a few bars of \ce{O2} and little to no steam remaining. Further constraints on the initial planet composition will require more stringent mass bounds and XUV flux measurements. 

Future observations of GJ1132b's atmosphere will allow us to probe these scenarios. The planet's transmission and emission spectra are sensitive to the relative abundances of O$_2$, H$_2$O, and other species \citep{Benneke2012}, and these spectra will be measurable with deep observations from the ground \citep{Snellen2013, Rodler2014} or from space \citep[][and references therein]{Cowan2015, Barstow2016}. Complementary JWST observations of GJ1132b's thermal phase curve could reveal its total atmospheric mass, and therefore determine whether the present-day atmosphere is thick or tenuous \citep{Koll2015,Selsis2011}. Detection of \ce{O2}-\ce{O2} collisionally-induced absorption features may also be used to constrain the presence and total pressure of a massive \ce{O2} atmosphere \citep{Schwieterman2016}.

Our model is applicable to a wide range of exoplanets inwards of their habitable zones. For instance, water loss from the recently discovered TRAPPIST-1 system \citep{Gillon2016} was modeled by \citet{Bolmont2016}. As the host star is an ultra-cool dwarf and the planets therefore receive less total XUV flux, they may retain both massive water vapor and \ce{O2} atmospheres, although conclusions must await both planet mass determinations, as well as detailed application of our model. Application to planets such as TRAPPIST-1d, which is potentially within the habitable zone, will require altering our atmospheric model to allow for condensible water vapor. In our own solar system, Venus may have experienced the loss of a similar steam-rich atmosphere as posited here for GJ 1132 b, but with nearly 10 times less stellar insolation, the escape rates from Venus should have been much lower and magma ocean cooling should have been much faster. Future application of this model to Venus may help confirm whether an early magma ocean could have taken up the \ce{O2} produced by atmospheric loss as suggested by \citet{Gillmann2009} and others. 
	
\section{Acknowledgements}
The authors thank Colin Goldblatt for providing the runaway greenhouse OLR data used to perform the intercomparison shown in Figure~\ref{fig:LBL_Goldblatt_intercomp}, and an anonymous referee for a positive and helpful review.
The line-by-line opacity computations in this paper were run on the Odyssey cluster supported by the FAS Division of Science, Research Computing Group at Harvard University.
LS and DS acknowledge support from the Simons Foundation. ZKBT gratefully acknowledges support from the MIT Torres Fellowship for Exoplanet Research.

\end{document}